\documentclass{aa}

\def\gtaprx {\lower .1ex\hbox{\rlap{\raise .6ex\hbox{\hskip .3ex
 {\ifmmode{\scriptscriptstyle >}\else {$\scriptscriptstyle >$}\fi}}}
 \kern -.4ex{\ifmmode{\scriptscriptstyle \sim}\else
 {$\scriptscriptstyle\sim$}\fi}}} 
\def\ltaprx {\lower .1ex\hbox{\rlap{\raise .6ex\hbox{\hskip .3ex
 {\ifmmode{\scriptscriptstyle <}\else {$\scriptscriptstyle <$}\fi}}}
 \kern -.4ex{\ifmmode{\scriptscriptstyle \sim}\else
 {$\scriptscriptstyle\sim$}\fi}}} 
\def\etal {et al. }
\def\littleprime{\ifmmode{\scriptscriptstyle \prime }
 \else{\hbox{$\scriptscriptstyle \prime$ }}\fi}
\def\littless{\ifmmode{\scriptscriptstyle s }
 \else{\hbox{$\scriptscriptstyle s $ }}\fi}
\def\littlemm{\ifmmode{\scriptscriptstyle m }
 \else{\hbox{$\scriptscriptstyle m $ }}\fi}
\def\littlehh{\ifmmode{\scriptscriptstyle h }
 \else{\hbox{$\scriptscriptstyle h $ }}\fi}
\def\littlecirc{\ifmmode{\scriptscriptstyle \circ }
    \else{\hbox{$\scriptscriptstyle \circ $ }}\fi}
\def\rasec{\raise .9ex \hbox{\littless}} 
\def\arcsec{\raise .9ex \hbox{\littleprime\hskip-3pt\littleprime\hskip-3pt}} 
\def\ramin{\raise .9ex \hbox{\littlemm}} 
\def\arcmin{\raise .9ex \hbox{\littleprime}}
\def\hrs{\raise .9ex \hbox{\littlehh}} 

\def\degree{\raise .9ex \hbox{\littlecirc}} 
\def\magpoint{\hbox to 2pt{}\rlap{\hskip -.5ex \arcmm}.\hbox to 2pt{}} 
\def\arcsspoint{\hbox to 1pt{}\rlap{\arcss}.\hbox to 2pt{}} 
\def\arcsecpoint{\hbox to 1pt{}\rlap{\arcsec}.\hbox to 2pt{}} 
\def\arcminpoint{\hbox to 1pt{}\rlap{\arcmin}.\hbox to 2pt{}} 
\def\degreepoint{\hbox to 1pt{}\rlap{\degree}.\hbox to 2pt{}}
\def\lax{{$\mathrel{\hbox{\rlap{\hbox{\lower4pt\hbox{$\sim$}}}\hbox{$<$}}}$}}
\def\gax{{$\mathrel{\hbox{\rlap{\hbox{\lower4pt\hbox{$\sim$}}}\hbox{$>$}}}$}}         

\usepackage{graphicx}
\usepackage{epsfig}
\usepackage{rotating}
     
\begin{document}

\newsavebox{\rotbox}


\title{Further Clues to the Nature of Composite LINER/H II Galaxies}

\author{Mercedes  E.  Filho \inst{1,2,3} \and 
Filippo Fraternali \inst{{\bf 1},4}   
\and 
Sera Markoff \footnote{NSF Astronomy \& Astrophysics
  Postdoctoral Fellow} \inst{5}
\and
Neil M. Nagar \inst{1}
\and
Peter  D. Barthel \inst{1}
\and 
Luis C. Ho \inst{6}
\and 
Feng Yuan \inst{7}
}


\institute{Kapteyn  Astronomical  Institute, P.O.~Box 800, 9700
  AV Groningen, The Netherlands
\and
Istituto di Radioastronomia, CNR, Via P. Gobetti, 101, 40129 Bologna, Italy 
\and
Centro de Astrof\'\i sica da
  Universidade do Porto, Rua das Estrelas, 4150 -- 762 Porto, Portugal 
\and 
ASTRON, P.O.~Box 2, 7990 AA Dwingeloo, The Netherlands
\and 
Massachusetts Institute of Technology, Center for Space Research, 
77 Massachusetts Av., R. NE80 -- 6035, Cambridge, Massachusetts 02139, U.S.A. 
\and
The Observatories of the Carnegie Institution of Washington, 813
  Santa Barbara Street, Pasadena, California 91101, U.S.A.
\and
Purdue University, Department of Physics, 525 Northwestern Av., West Lafayette, Indiana, 47907 -- 2036, U.S.A.}   

\date{Received XX; accepted XX}


\abstract  {  We  have  analyzed  new,  archival  and  published  high
resolution  radio and  X-ray  observations of  a  sample of  composite
LINER/H~{\sc ii} galaxies known  to exhibit AGN-like properties.  Five
of the 16 AGN  candidates have milliarcsecond-scale detections and are
found to display a compact, flat spectrum, high brightness temperature
radio core, four of which  also exhibit extended radio emission.  Five
of  the eight  AGN  candidates with  available  high resolution  X-ray
observations were found to possess a hard X-ray nuclear source, two of
which  have  no  milliarcsecond  radio detection.  The  combined  high
resolution radio  and X-ray  data yield a  50\% detection rate  of low
luminosity AGN among the AGN  candidates, which translates into a 12\%
detection rate  for the entire composite LINER/H~{\sc  ii} sample.  In
the  sources  where  the  AGN  has been  unambiguously  detected,  the
ionizing power of  the AGN is not sufficient  to generate the observed
emission  lines, unless  the  hard X-rays  are  heavily obscured.   We
attempt to apply a canonical advection-dominated accretion flow (ADAF)
and jet model  to the sample sources in order  to explain the observed
radio  and X-ray  emission. While  ADAFs  may be  responsible for  the
observed emission  in submillijansky radio cores  like NGC\,7331, they
do  not appear  consistent with  the  radio emission  observed in  the
milliarcsecond-scale radio detected cores; the latter sources are more
likely  to  have  an   energetically  important  contribution  from  a
radio-emitting jet.


\keywords{galaxies: active --- galaxies: nuclei --- radiation mechanisms: general}

}


\titlerunning  {Further Clues to  the Nature  of Composite  LINER/H II
Galaxies} \authorrunning{Filho \etal}

\maketitle


\newpage

\section{Introduction}

The radio  emission observed in  nearby galaxies can be  attributed to
stellar   and  interstellar   processes  --   non-thermal  synchrotron
radiation  from electrons accelerated  by supernova remnants
(SNRs)  and thermal  emission from  H~{\sc ii}  regions --  or  to the
presence  of accreting  black holes.   In the  former case,  the radio
emission   coincides   with   the   infrared  and   optical
emission-line sources, which are themselves tracers of young stars and
supernovae.   The radio  emission originates  in diffuse,  low surface
brightness regions that do not extend beyond the stellar light distribution.  On the
other  hand,  radio sources  produced  by  accreting  black holes  are
generally  characterized by  compact, nuclear  radio  cores, sometimes
associated  with  jets  and  lobes. In  many  instances  circumnuclear
starburst  regions  and  black  holes  coexist  spatially,  making  it
difficult to disentangle the two  components in the optical regime and
also in the radio.  Furthermore, at high resolution, even star-forming
galaxies  may  show  compact  radio  sources,  associated  with  radio
supernovae (RSN)  and SNRs  (e.g. Weiler \etal  1986). In an  active
galactic nucleus (AGN), the
presence of a flat radio spectrum source ($\alpha$\lax0.5;
F$_{\nu}\propto\nu^{-\alpha}$)   
indicates  synchrotron  self-absorption  in
components whose brightness temperatures are comparable to the kinetic
temperatures (mc$^2$/k; T$_{\rm  B}$\gax10$^5$~K) of the relativistic
electrons  producing   the  synchrotron  radiation.    Steep  spectrum
emission in star-forming galaxies ($\alpha$\gax0.5) arises from the
non-thermal emission of SNRs, whereas flat spectra are consistent with
free-free  absorption  of  optically  thin synchrotron  radiation  and/or
thermal  emission  (T$_{\rm  B}$\lax10$^4$~K)  originating  from  the
electrons in  H~{\sc ii}  regions.  So, while  the spectra  of thermal
emitters may  mimic those of  partially opaque synchrotron  sources, a
combination of brightness temperature  and radio spectral index of the
strong nuclear  radio components can indicate whether  the emission is
thermal  (T$_{\rm   B}$\lax10$^4$~K;  Condon  1992)   or  non-thermal
(T$_{\rm    B}$$>$10$^5$~K)   and    whether   optically    thin   thermal
($\alpha\sim$0.1),   self-absorbed   synchrotron  ($\alpha$$<$0.5),   or
optically thin synchrotron ($\alpha$$\sim$0.5--0.8) radiation dominates.

We have therefore  applied this radio technique to  a sample of mildly
active  galaxies -- composite  LINER/H~{\sc ii}  galaxies --  known to
exhibit  both  starburst  and  AGN characteristics  in  their  optical
spectra.  The  motivation for studying  this class of galaxies  is to
study the demographics  of black hole accretion in  cases suspected of
weak AGN  activity. This is  an important issue because,  if composite
sources  do  indeed  harbour  a  genuine active  nucleus,  then  their
considerable  number (13\% of emission-line nuclei with B$_{\rm T}<$ 12.5~ mag; Ho,  Filippenko \&  Sargent 1997)  will 
imply a significant contribution to the AGN population.

\section{Sample Selection} 

In  Filho, Barthel \& Ho (2000, 2002)  we  have  presented  a   study  of  the  radio
characteristics  of a  complete sample  of composite  LINER/H~{\sc ii}
galaxies, i.e., galaxies that  are hypothesized to be LINERs spatially
contaminated  by circumnuclear  star-forming regions  (Ho  1996).  The
first part of our project consisted in gathering arcsecond-scale radio
information (published  or new) of the 65  composite sources contained
in the  magnitude-limited Palomar  survey (Ho, Filippenko \& Sargent
1995, 1997a).   Their radio  properties
(Filho, Barthel \& Ho 2000, 2002) indicate composite sources come in
two  types: AGN-like  sources  with compact  cores and  starburst-like
sources dominated  by diffuse emission, co-spatial  with the projected
galactic  disk.  The  complete  composite source  sample contained  14
($\sim$25\%)  AGN  candidates as  revealed  by compact,  flat/inverted
radio  spectrum cores on  arcsecond scales,  with peak  flux densities
above 1~mJy.   NGC\,660 and  NGC\,7331 have been  included in  the AGN
candidate sample  although they have sub-mJy radio  peaks as explained
in Filho, Barthel \& Ho (2002).  These sixteen  sources comprise the
present sample of candidate AGN-driven composite LINER/H~{\sc ii} galaxies.

\section{New and Published Radio Observations}

Radio imaging at sub-arcsec and mas resolution of several of the sample
sources  have already been published.  Our observations,  using the  Very Large
Array (VLA) and the Very Large Baseline Array (VLBA), complement these
observations and yield a complete picture of the occurrence of compact
nuclei in the sample.

In  this section  we  report  on high  resolution  VLA 
observations  of  12  (four new and eight published)  and 
multiwavelength VLBA observations of 15 (13 new and two published) out
of  the 16  sample sources.  Regarding the former, of the possible eight galaxies 
with no published sub-arcsec-scale radio information,  four were chosen to fit
the observational window assigned to us by the VLA. 
As for  the latter,  NGC\,7331 was not observed with the VLBA  
 because    its     radio    core    flux    density    at
1\arcsecpoint5~resolution is  well below 1~mJy at both  1.4 and 5\,GHz
(Cowan, Romanishin \& Branch 1994) and NGC\,5866 and NGC\,4552 because
they had been previously observed and detected with the VLBA by Falcke
\etal  (2000)  and  Nagar  \etal  (2002),  respectively.  Following  a
description of our observations, we will combine the new and published
data in a coherent analysis of the sample.

\subsection{Observations and Data Reduction}

VLA observations of four sources  in the sample were performed on 1999
September 5  with the A-array,  X-band system (8.4\,GHz).  Two  IFs of
50\,MHz each were combined to  give a total bandwidth of 100\,MHz.  In
its  A-array  configuration, the  VLA  yields  typical resolutions  of
0\arcsecpoint25 at 8.4\,GHz (which  corresponds to 12~pc at D=10~Mpc).
Observations of the sample sources were interspersed with observations
of nearby phase calibrators, for  a total integration time of about 10
minutes  per   galaxy.   The  primary  flux   calibrator  was  3C\,286
(1328+307)  with  adopted 8.4\,GHz  flux  densities  of 5.1915~Jy  and
5.1702~Jy for  IF1 and  IF2, respectively. The  uncertainty associated
with the flux  calibration procedure is mainly due  to the uncertainty
in the absolute  flux density of 3C\,286, which  is conservatively set
to 5\%.  Antenna gains were found to behave well throughout the observations. 
  
VLBA observations  of 13 out of  the 16 sample galaxies  were obtained in
five sessions --  2000 June 22, 2001 September  1, 2001 September 8,
2001 September  17 and 2001 October  6 -- with the  standard ten VLBA
antennas. The standard observing  frequency was 5\,GHz, although some
sources  have  multi-wavelength  observations.   The  four  8\,MHz  IFs
yielded  a total bandwidth  of 32\,MHz.   Typical resolutions  for the
VLBA at  5\,GHz are 2~mas  (which corresponds to 0.1~pc at D=10~Mpc).  
Phase  referencing was
performed,  which includes several  minute on-source  scans alternated
with scans of nearby phase calibrators, yielding generally 180 (in some
cases 45) minutes total integration time on each target source.

Reduction of both  the VLA and VLBA data  was performed using standard
NRAO AIPS  (version 15Oct99) image processing  routines. After initial
deleting of bad data points  and calibration, the data were `cleaned'.
The AIPS `cleaning'  task IMAGR was employed to  Fourier transform the
data  and  remove  sidelobes  from  the maps  in  an  interactive  and
iterative  mode.   Some sources were observed in multiple VLBA runs
separated by several days. For these sources, maps were made for
each run separately, since combining the data did not result in a 
better map.
Full resolution  images,  having a  0\arcsecpoint25
synthesized beam, were obtained for the VLA data and 2~mas synthesized
beam at 5\,GHz  for the VLBA data.  The image  noise level reached the
theoretical level  of $\sim$0.07~mJy/beam (uniform  weighting; Perley,
Schwab \& Bridle 1989) to within a factor of two for the VLA data and
$\sim$0.05~mJy/beam  for the  VLBA data.   Phase  self-calibration was
employed on  the strongest sources,  leading to some reduction  in the
image  noise   level.  The  8.4\,GHz  VLA calibrated data provided by
Dr. Jim Ulvestad for NGC\,660  have  been
re-analysed,  yielding  slightly  different radio  parameters than
those presented in Filho, Barthel \& Ho (2002).

The image  noise levels were measured  with the AIPS task  IMSTAT in a
source-free region.  Using the AIPS  task IMFIT, the  brightness peaks
of the radio sources were fitted with single bi-dimensional Gaussians.
 
\subsection {Results of the Radio Observations}

Four of four VLA and three of 13 newly observed VLBA sample
sources were detected. Table~1 at the end of the paper lists 
image parameters for the new data, where we have also included relevant published radio
data from Nagar \etal (2000, 2002), Cowan, Romanishin \& Branch (1994; NGC\,7331)
and  Falcke \etal (2000; NGC\,5866).   We refer  to Filho,  Barthel \&  Ho (2000,
2002)  for an  extensive discussion  on  the radio  properties of  the
sample sources.   Undetected sources  have been given  5$\sigma$ upper
limits  for  the  VLA and  VLBA  data.   We  estimate the  radio  peak
positions to be accurate  to within \lax0\arcsecpoint2~for the VLA and
\lax1~mas  for the  VLBA data,  due to  the good  phase  solutions and
accurate positions  of the phase calibrators.   Sources are considered
unresolved if  their deconvolved source  sizes are less than  half the
beamwidth in any component.

Fig. 1--2 at the end of the paper show the highest signal-to-noise 
radio maps of the newly detected VLBA sources. The contour levels are 
the following multiples of the $rms$
noise (Table~1)  in the map: $-$3,  $-$2.1, 2.1, 3, 4.2,  6, 8.4, 12,
16.8, 24.

\subsubsection{Radio Core Properties}

In Table~2  we list the  radio properties of the  detected mas-scale
sample sources (from new or published data). Unless otherwise  mentioned, we have used the longest
integration time  VLBA 5\,GHz radio  flux densities to calculate  the spectral
indices  and brightness  temperatures.  Radio  spectral indices
should be taken  with caution, due to non-simultaneity  and mismatch in
resolution.   For the multiple radio components  in NGC\,5846,  we  have given
5$\sigma$  limits to  any 5\,GHz  emission  in order  to obtain  upper
limits  for  the  spectral   indices.   We  calculate  the  brightness
temperature as:

\begin{center}

T$_{\rm B}$=7.8$\times$10$^6\left(\frac{\rm F_{\nu}}{\rm
    mJy}\right)\left(\frac{\theta}{2.5\,\rm mas}\right)^{-2}\left(\frac{\nu}{5\,\rm GHz}\right)^{-2}$

\end{center}  

\noindent where  F$_{\nu}$ is the peak  flux density at  5\,GHz in the
VLBA maps (Table~1)  and $\theta$ is the FWHM  of the Gaussian beam,
taken to be typically 2.5~mas at 5\,GHz. For unresolved sources, the
quoted brightness temperatures should be considered lower limits.


\setcounter{table}{1}

\begin{table*}[!ht]

\footnotesize

\begin{center}

\begin{minipage}{106mm}

\caption{{\bf Nuclear radio properties (published or new).} 
Col. 1: Source name.
Col. 2: High resolution radio flux density.
Col. 3: Observing frequencies.
Col. 4: Radio spectral index.
Col. 5: Radio variability.
Col. 6: Peak brightness temperature.
Col. 7: Note.
}

\begin{tabular}{l c c cc l c c}

\hline
\hline

  & F$_{\rm radio}$ & $\nu$ & 
\multicolumn{2}{c}{} &  & T$_{\rm B}$ &  \\
Galaxy & (mJy) & (GHz) &
\multicolumn{2}{c}{$\alpha$} &  var. & (K) & Note \\
 (1) & (2) & (3) & \multicolumn{2}{c}{(4)}  &
(5)   & (6) & (7) \\

\hline

NGC\,524  & 1.95/1.5 & 8.4/5 & \multicolumn{2}{r}{$>-$0.5} & \ldots & 7.0$\times$10$^{6}$ & a \\

NGC\,4552 & 102.1/99.5  & 8.4/5    &  \multicolumn{2}{r}{$\sim$0.0}    & yes     & 7.8$\times$10$^{8}$ & b,c \\

NGC\,5354 & 8.6/8.7 & 5/2.3  & \multicolumn{2}{r}{0.0}    & yes?    & 1.9$\times$10$^{7}$ & \ldots  \\

NGC\,5846A       & 1.5/2.8  & 5/2.3 & \multicolumn{2}{r}{0.8}     & yes?    & 4.7$\times$10$^{6}$ &\ldots \\

NGC\,5846B       & 0.6/$<$0.8  & 5/2.3  & \multicolumn{2}{r}{$>$0.4} & \ldots & 2.3$\times$10$^{6}$ & \ldots \\

NGC\,5846C       & 0.7/$<$0.8  & 5/2.3  & \multicolumn{2}{r}{$>$0.2} & \ldots & 2.3$\times$10$^{6}$ & \ldots \\  


NGC\,5846D       & 1.9/$<$0.3 & 15/5  & \multicolumn{2}{r}{$>-$1.8}  & yes? & 7.6$\times$10$^{6}$ & d \\  

NGC\,5866 & 7.5/8.4 & 15/5   & \multicolumn{2}{r}{0.1}    & \ldots & 1.0$\times$10$^{8}$ & e \\

NGC\,7331 & 0.121/0.234 & 5/1.5 & \multicolumn{2}{r}{0.5} & \ldots & \ldots & f  \\  

\hline

\end{tabular}

\smallskip

{\sc Notes} -- (a)  spectral  index  calculated  using  the  8.4\,GHz,
2\arcsecpoint5~resolution peak flux density (Filho, Barthel \& Ho 2002)
and VLBA 5\,GHz integrated flux density (this paper); (b) spectral index calculated using
the  8.4\,GHz,   2\arcsecpoint5~resolution peak flux density  (Filho,
Barthel \& Ho 2002)  and VLBA  5\,GHz integrated flux density  (Nagar \etal  2002); (c)
brightness  temperature calculated  using  the VLBA  5\,GHz peak flux density 
(Nagar \etal 2002); (d) brightness temperature calculated using the VLBA
15\,GHz  peak flux density (this paper);  (e) spectral  index calculated  using the VLA,
15\,GHz,  0.15\arcsec~resolution (Nagar  \etal 2002)  and  VLBA 5\,GHz
integrated 
flux density (Falcke \etal 2000); (f) spectral index calculated using the VLA,
1.5  and   5\,GHz,  2\arcsec~resolution integrated flux density (Cowan,
Romanishin \& Branch 1994).

\end{minipage}

\end{center}

\end{table*}

\normalsize


In a  distance-limited sample of  low luminosity Seyferts,  LINERs and
composite  LINER/H~{\sc ii}  galaxies  imaged with  the  VLA (150  mas
resolution),  50\% of  the sources  showed flat  spectrum  radio cores
(Nagar  \etal 2002, 2000).  Subsequent VLBA  5\,GHz observations  of a
flux-limited  subsample (16  sources with flux densities $>$3 mJy)  yielded  a 100\%
detection rate  of low  luminosity AGN (LLAGN;  Nagar \etal  2002; see
also  Falcke \etal  2000). In  comparison, the  present sample  of AGN
candidates yielded a VLA radio core  detection rate of 83\% (ten of
12  sources with VLA data, Table~1).   Follow-up VLBA observations
(including  NGC\,4552; Nagar  \etal 2002  and NGC\,5866;  Falcke \etal
2000) have  revealed that  33\%  (five of  15) of  the sources  have
mas-scale radio cores. These results translate to an overall radio detection rate of 8\%
(five of 65) among the entire composite  LINER/H~{\sc ii} sample. 
The  mas-scale non-detections were sources with
VLA flux densities  $<$3 mJy;  thus, our results  are consistent with  those of
Falcke \etal  (2000) and  Nagar \etal (2002).   Moreover, as  has been
pointed out in these papers  and discussed also in Filho \etal (2002),
scaling  arguments suggest  that  the radio  non-detections could well be  lower
power versions of the radio  detections and thus harder to detect. The
true AGN fraction in composite  LINER/H~{\sc ii} galaxies is likely to
be higher.

Of the  four VLA observed  sample sources, only NGC\,660  was slightly
resolved at VLA A-array resolution.  The present new and published VLBA data show unresolved
cores in  NGC\,524 (Fig. 1{\it  a}), NGC\,4552 (Nagar \etal  2002) and
NGC\,5354 (Fig. 1{\it b,c}).  Some sources also show evidence of
radio variability, as seems the  case for NGC\,4552 (see discussion in
Filho,  Barthel \&  Ho 2000).  Furthermore, all  VLBA  detected sample
sources (including NGC\,4552; Nagar \etal 2002 and NGC\,5866; Falcke \etal 2000) 
show flat spectrum radio cores, although the inverted
spectrum  in NGC\,524  is most  probably  an  effect of  resolution
mismatch.  The origin  of the mas-scale radio emission  is unlikely to
be free-free emission.  The detected radio cores show  brightness temperatures
several orders of magnitude  above the thermal limit (10$^4$~K; Condon
1992). Moreover,  as argued in  Falcke \etal (2001), a  thermal origin
for the radio emission would imply much higher X-ray fluxes than those
observed  (see Section~4).  Therefore, the  combination  of brightness
temperatures   in   excess    of   10$^7$~K   and   spectral   indices
$\alpha$\lax0.5 shows  that the radio emission can  be identified with
(multi-component)  non-thermal,   self-absorbed  synchrotron  emission
related to an accreting black hole.

On the other hand, the  nature of the radio components associated with
NGC\,5846 is more complex. Our 2.3, 5, and 15\,GHz VLBA maps of NGC\,5846 show
a multiple radio source roughly
aligned  in  the North-South  direction (Fig. 2{\it a,b,c}) -- components NGC\,5846A, NGC\,5846B, 
NGC\,5846C and NGC\,5846D. Apparently we have resolved the single core observed with the VLA
A-array      (0\arcsecpoint25~resolution)     and      VLA     C-array
(2\arcsecpoint5~resolution; Filho, Barthel \& Ho 2002). At 5\,GHz, 11~mas separate  the
Southernmost (component  A) from the  Northern ones (components  B and
C).  Component A
appears in both the 2.3\,GHz  and 5\,GHz maps (Fig 2.{\it a,b}), remaining unresolved at
the latter frequency. However, it  is not detected at 15\,GHz (Fig 2.{\it c}), showing
the  source to  be steep  spectrum and/or variable. 
The  steepness of  the spectrum  and  high brightness
temperature (Table~2)  make it a  candidate RSN (e.g.   Tarchi \etal
2000). However, the compactness  of the  source,
associated with the high brightness temperature and given its linear radio structure,
suggests that component  A is  the variable radio core of NGC\,5846.

\subsubsection{Extended Radio Nuclear Components}

There are three (NGC\,5354, NGC\,5846 and also NGC\,4552; Nagar \etal 2002), 
perhaps four (see discussion on NGC\,5866 in
Falcke \etal 2000) sources in the sample which show extended structure on mas-scales.

Although unresolved  on arcsec-scales  (Filho, Barthel \& Ho 2000;  Nagar \etal
2000), and  in our  VLBA maps,  a recent high signal-to-noise VLBA image of
NGC\,4552 (Nagar  \etal 2002) revealed a core  plus two-sided emission
in the East and West direction.

Our 0\arcsecpoint2~resolution VLA map of NGC\,5354 shows an unresolved
core, whereas  our longer  integration VLBA map  of this  source shows
about 4~mJy in  symmetric double-sided emission to the  East and West
(Fig. 1{\it c}),
suggesting the presence  of jets cradling an AGN  core.  This extended
emission  is not  seen, however,  on our other VLBA  shorter exposure
maps, most probably due to low signal-to-noise.

Our first epoch 5\,GHz VLBA map shows a triple source which we have designated as components 
NGC\,5846A, B, and C. Furthermore, at 15\,GHz we detect another component 
(NGC\,5846D; Fig. 2{\it c}), about \lax60~mas from the 5\,GHz radio position of NGC\,5846A and 
which cannot be identified with any of the components in the 5\,GHz map. 
NGC\,5846A has been discussed in the previous subsection, but
the nature of the latter three
components is more difficult to constrain. 
Components B, C and  D appear only at one frequency each,
and  most  probably  have an  inverted 
(NGC\,5846D)or steep spectrum (NGC\,5846B  and  NGC\,5846C), 
suggestive of optically thin synchrotron emission. The apparent inverted 
nature of the spectrum of NGC\,5846D could also be due to 
radio variability. All three  components also appear  to have  a somewhat
amorphous, resolved   morphology    and    high     brightness    temperatures
typical  for
non-thermal sources (Table~2). Furthermore, they are more or less aligned with component
A in the North-South direction.  One scenario is that these components
are  compact SNRs near  the center  of the  galaxy (e.g.  Tarchi \etal
2000) or they may be emission associated with a radio jet.

\section{Archival and Published X-ray Observations}

For most of our sample sources, published X-ray data are
available -- they  mainly refer to low resolution observations  
not suited to our  purposes. Therefore, in this section we present the results of
the analysis of archival and/or published high
resolution X-ray {\it
  Chandra} data for the sample sources. 

\subsection{Observations and Data Reduction}

Public  {\it Chandra}  ACIS-S archival data are  available  for seven of the sample
sources, three of which have already been published (NGC\,1055 and
NGC\,3627; Ho \etal 2001 and NGC\,5846; Trinchieri \& Goudfrooij
2002). NGC\,5866  {\it Chandra} data have been published (Terashima \& Wilson
2003) but are not as yet publicly available.

For reasons of self-consistency and homogeneity, all publicly
available {\it Chandra} data were reduced and analyzed using  the CIAO (version
2.2) and  XSPEC (version 11.2)  packages.  We removed  high background
time by  excluding events exceeding $\pm$~3$\sigma$ of  the mean image
count rate. The energy range  was restricted to the 0.4$-$10 keV band.
We searched for X-ray nuclei  coincident with the radio core positions
(Filho, Barthel \& Ho 2000, 2002; Nagar  \etal 2000, 2002; Section~3) to within
the  positional accuracy  of  {\it Chandra}  ($\sim$ 1\arcsec).   When
found, spectra were  extracted from a circular region  with a diameter
of 2.5\arcsec~($\sim$5 pixels on the ACIS image) in order to spatially
match both the VLA C-array radio maps (Filho, Barthel \& Ho 2000, 2002) and the
regions from where the  H$\alpha$ luminosities were obtained (Ho, Filippenko \& Sargent
1997).  The background was extracted from an annulus around the source
with an inner diameter of 5\arcsec.

Possible  difficulties in pinpointing  a hard  X-ray nucleus  can arise
from  the  presence  of   ultraluminous  X-ray  sources  (ULXs).  Such
confusion   is   minimized   by    the   high   resolution   of   {\it
Chandra}. However,  this possibility can  not be totally  excluded for
galaxies like NGC\,5846 (see Section~4.2), where several X-ray sources
are  detected in  the nuclear  region. With  the possible  exception of
NGC\,4552, pile-up  effects  should not be a problem, given  the relatively
low count rates.  Even in the case of  NGC\,4552, a pile-up correction
with  a subsequent factor  2 difference  in flux  would not  alter the
following results.

For three of the X-ray sources (NGC\,4552,  NGC\,5846, NGC\,7331), the longer exposure times
($\tau\simeq$30~ks), allowed 
 a  spectral analysis. 
We considered a power law model plus, for NGC\,4552, a thermal plasma
component  (Raymond \& Smith  1977).  The  column densities  have been
fixed  to the  Galactic  values  (Dickey \&  Lockman  1990).  The fitted
photon indices (1.5  -- 2.3) are consistent with  values obtained for
LLAGN (Terashima \& Wilson 2003).
For sources observed with shorter exposure times, the  count rates
were converted  to X-ray fluxes  assuming a simple  power-law spectrum
with a photon index of  1.7 and Galactic absorption.  Upper limits for
non-detections  were  calculated  after  running  a  source  detection
procedure ($wavdetect$)  on the  full resolution {\it  Chandra} image,
with  a  threshold  of  10$^{-6}$  (4.7$\sigma$  level).   Assuming  a
power-law  ($\Gamma$=1.7) and  Galactic absorption,  the  upper limits
were calculated to be equal to the flux of the weakest detected source
on  the image.   Our results are consistent  with  those obtained  by Ho \etal
(2001) for NGC\,1055 and NGC\,3627 and Trinchieri \& Goudfrooij (2002) for  NGC\,5846 (see Section~4.2).

\subsection{Results of the X-ray Observations}

The X-ray parameters as given by {\it Chandra} archival or published data 
are specified in Table~3.

Five of eight (63\%) sample sources with available {\it Chandra} data
were detected as having hard X-ray nuclear cores. Given the hypothetical weakness of the hard X-ray core, 
the three non-detections are most
likely related to the low integration time compared to the observations of the other sources; these have $\tau$ = 1--2 ks (Table~3).

Below we give  a brief description of the X-ray  properties of the AGN
candidates.


\begin{table*}[!ht]

\setcounter{table}{2}

\footnotesize

\begin{center}

\begin{minipage}{142mm}

\caption{{\bf X-ray Properties (published or archival).}
Col. 1: Source name.
Col. 2: Exposure time.
Col. 3: Photon index (and Raymond-Smith plasma temperature) of the
power law fit and assumed. 
Col. 4: Unabsorbed X-ray flux in the 0.5--2~keV band.
Col. 5: Unabsorbed X-ray flux in the 2--10~keV band.
Col. 6: Logarithm of the soft X-ray luminosity (0.5--2~keV).
Col. 7: Logarithm of the hard X-ray luminosity (2--10~keV).
Col. 8: Note.
}

\begin{tabular}[ht!]{l c cc rc rc cc cc c}

\hline
\hline

 & $\tau$ & Model & \multicolumn {2}{c}{F$_{\rm soft}$} &
 \multicolumn {2}{c}{F$_{\rm hard}$} & \multicolumn {2}{c}{log L$_{\rm
     soft}$} & \multicolumn {2}{c}{log L$_{\rm hard}$} &  \\
Galaxy & (ks) & $\Gamma$ (+kT) & \multicolumn {2}{c}{(erg cm$^{-2}$
s$^{-1}$)} & \multicolumn {2}{c}{(erg cm$^{-2}$ s$^{-1}$)} & 
\multicolumn {2}{c}{(erg s$^{-1}$)} & \multicolumn {2}{c}{(erg
  s$^{-1}$)} & Note \\
 (1) & (2) & (3) & \multicolumn {2}{c}{(4)} & \multicolumn
{2}{c}{(5)} & \multicolumn
{2}{c}{(6)} & \multicolumn
{2}{c}{(7)} & (8) \\

\hline

N\,660  &  1.92  & 1.7 & \multicolumn
{2}{r}{7.12$\times$10$^{-15}$}    & \multicolumn
{2}{r}{1.30$\times$10$^{-14}$} &  \multicolumn
{2}{r}{38.07}   &  \multicolumn
{2}{r}{38.33}  & a \\

N\,1055 &  1.14  & 1.7 & \multicolumn {2}{r}{$<$7.11$\times$10$^{-15}$} & \multicolumn {2}{r}{$<$1.30$\times$10$^{-14}$} & \multicolumn
{2}{r}{$<$38.13} & \multicolumn
{2}{r}{$<$38.39} & a,b \\

N\,3245  & 9.63 &  1.7 &  \multicolumn
{2}{r}{7.43$\times$10$^{-15}$}    & \multicolumn
{2}{r}{1.36$\times$10$^{-14}$}  &  \multicolumn
{2}{r}{38.63} &  \multicolumn
{2}{r}{38.89} & a \\

N\,3627 &  1.75  & 1.7 & \multicolumn {2}{r}{\lax1.20$\times$10$^{-14}$} & \multicolumn {2}{r}{\lax2.19$\times$10$^{-14}$} & \multicolumn
{2}{r}{\lax37.79} & \multicolumn
{2}{r}{\lax38.06} & a,b \\

N\,4552 & 54.42 & 1.51 (0.95) & \multicolumn
{2}{r}{4.81$\times$10$^{-14}$}    & \multicolumn
{2}{r}{7.58$\times$10$^{-14}$}   &  \multicolumn
{2}{r}{39.21}   &  \multicolumn
{2}{r}{39.41} & c  \\

N\,5846 &  29.86 & 2.29  & \multicolumn
{2}{r}{3.35$\times$10$^{-15}$} & \multicolumn
{2}{r}{2.44$\times$10$^{-15}$} &  \multicolumn
{2}{r}{38.51} &  \multicolumn
{2}{r}{38.37} & d,e \\

N\,5866 & 2.25 & 2.0  &  \multicolumn
{2}{c}{\ldots} & \multicolumn {2}{r}{$<$6.37$\times$10$^{-15}$} &
 \multicolumn
{2}{r}{\ldots} & \multicolumn
{2}{r}{$<$38.26} & a,f \\

N\,7331 &  29.46 & 1.46     & \multicolumn
{2}{r}{4.46$\times$10$^{-15}$} & \multicolumn
{2}{r}{1.16$\times$10$^{-14}$}     &  \multicolumn
{2}{r}{38.04} & \multicolumn
{2}{r}{38.45} & g \\

\hline

\end{tabular}

{\sc Notes} -- (a) Photon index fixed to the respective value; (b)
also in Ho \etal (2001); (c) error in the photon index is
$\pm$0.09; (d) also in Trinchieri \& Goudfrooij (2002); (e) error in
the photon index is $\pm$0.38; 
(f) quoted {\it Chandra} ACIS-S data from Terashima \& Wilson (2003); (g) error in the photon index is $\pm$0.32.

\end{minipage}

\end{center}

\end{table*}

\normalsize

\smallskip

{\it  NGC\,660}  The soft X-ray region
is   about  0\arcminpoint5  in   extent  in   the  Northeast-Southwest
direction, about the size and orientation of the radio emitting region
(see Filho, Barthel \& Ho 2002). The  hard X-ray  nuclear region,  spatially coincident
with  the radio  core  (Filho, Barthel \& Ho 2002),  comprises  only about  10
counts, such that no spectral fit was possible.

\smallskip

{\it NGC\,1055} No  obvious X-ray source -- hard or  soft -- was found
on the  {\it Chandra}  image (see also Ho \etal 2001). 
A source  detection procedure ($wavdetect$) was  run in
order to obtain an upper limit to the X-ray emission.

\smallskip

{\it NGC\,3245} There is a hard nuclear X-ray source coincident with the
optical nucleus. The soft X-ray emission is resolved and extended ($\sim$
6\arcsec) in the North-South direction.

\smallskip

{\it NGC\,3627} The soft X-ray emission extends over about 2\arcmin~in
a Northwest-Southeast direction, similar to the extent and orientation
of the triple-radio source (Filho, Barthel \& Ho 2000). No obvious nuclear X-ray source -- hard or soft -- was
found  on the  {\it  Chandra} image (see also Ho \etal 2001).   Although  the source  detection
procedure  ($wavdetect$) found a  source positionally  coincident with
the VLA  core (Filho, Barthel \& Ho 2000; Nagar \etal  2000, 2002), it  is the
weakest source  on the  image and  we prefer to  consider it  an upper
limit (see also Ho \etal 2001).

\smallskip

{\it NGC\,4552} The  soft X-ray emission extends almost  to the limits
of  the   optical  galaxy  ($\sim$2\arcmin),   exhibiting  a  somewhat
hourglass  structure.  We  find a  strong hard  X-ray  nuclear source,
coincident with the radio core position (Nagar
\etal 2002; Filho, Barthel \& Ho 2000).  The spectrum  of the nuclear source (1200 net
counts) yields a satisfactory fit  with a two component model -- power
law ($\Gamma$=1.51) plus Raymond-Smith thermal plasma (kT=0.95 keV).

\smallskip

{\it NGC\,5846} The  X-ray morphology of this source  is complex
(see  also Trinchieri \&  Goudfrooij 2002),  extending over  more than
3\arcmin, almost to the limits  of the optical galaxy. It exhibits two
`spiral  arms', directed to  the Sortheast  and Southwest.
There is a  weak hard X-ray nucleus, coincident with the  VLA radio core in
NGC\,5846, whose spectrum was fit with a power law ($\Gamma$=2.29).
The source we have identified with the nucleus coincides with the third
most luminous X-ray source in NGC\,5846, listed as source number 12 in Trinchieri \&
Goudfrooij (2002). Their quoted flux for this source is six times
higher then our estimation because they convert counts to fluxes
assuming a constant conversion factor. If we apply the same formulation, our
results are consistent with theirs by a factor of 1.4. 

\smallskip

{\it  NGC\,5866} Recent published {\it Chandra} observations  of this source  failed to
detect any hard  X-ray emission from the nucleus  (Terashima \& Wilson
2003).

\smallskip

{\it  NGC\,7331} The  {\it Chandra}  soft X-ray  image of  this galaxy
shows several blotchy regions of  emission, along about 3\arcmin, in a
North-South direction, similar to the morphology of the optical galaxy
and also the  radio (Cowan, Romanishin \& Branch 1994). Previous
observations with
ROSAT (Stockdale, Romanishin \& Cowan 1998) have detected a luminous
X-ray nucleus (few 10$^{40}$ erg s$^{-1}$). The {\it Chandra}
observation has resolved the ROSAT emission into several X-ray sources and revealed
a hard X-ray nucleus
positionally  coincident  with  the  radio  core  detected  in  Cowan,
Romanishin \& Branch  (1994).  The spectrum of the  nuclear source is
fit with a one component power law ($\Gamma$=1.46) model and the luminosity
is two  magnitudes lower ($\sim$10$^{38}$ erg s$^{-1}$) than the ROSAT value.

\section{Radio, Optical and X-ray Relations}

In Table~4,  at the  end of the paper, we summarize  the relevant
properties for  the sample sources from published, archival or new observations.  

With  the exception  of NGC\,3245  (Barth  \etal 2001),  there are  no
reliable dynamical black hole mass estimates  for the sample sources.  The next
best  method for  determining  black hole  mass  utilizes the  central
velocity  dispersion.  The  correlation  between black  hole mass  and
central  velocity  dispersion is tight (Ferrarese  \&
Merritt 2000; Gebhardt \etal 2000; Tremaine \etal 2002):

\begin{center}

$\rm  log \left(\frac {\rm  M_{\rm BH}}{\rm  M_\odot}\right) =  8.13 +
4.02 \; log \left[ \frac{\sigma_e}{200\,{\rm km s^{-1}}} \right]$

\end{center}

\noindent where $\sigma_e$ is the luminosity-weighted stellar velocity
dispersion measured at the effective radius.  However, in this case we
will simply use the value $\sigma_c$, the central velocity dispersion,
since  Gebhardt \etal (2000)  have shown  that the  difference between
$\sigma_e$ and $\sigma_c$ is only $\sim$10\%.

As can  be seen in Table~4,  the dynamical black  hole mass estimate
for NGC\,3245  (2$\times$10$^8$ M$_{\odot}$;  Barth \etal 2001)  is in
excellent agreement with the velocity dispersion estimate.

Nuclear B-band  magnitudes have been calculated  using the correlation
found for low luminosity  Seyfert~1s between M$^{\rm nuc}_{\rm B}$ and
the H$\beta$ (narrow plus broad) line luminosity (Ho \& Peng 2001):

\begin{center}

log (L$_{\rm H\beta}) = -0.34 \, {\rm M^{\rm nuc}_{\rm B}} + 35.1$

\end{center}

\noindent With the exception of the Type~1.9 (broad  H$\alpha$
emission present)  source NGC\,1161,  all the  sample sources  are  Type~2 (no
broad-line emission).  Therefore, we have  used only the  {\it narrow}
H$\beta$  line  luminosities  taken  from Ho,  Filippenko  \&  Sargent
(1997a) to estimate M$^{\rm nuc}_{\rm B}$. In the  case  of composite
galaxies,  the derived  M$^{\rm nuc}_{\rm  B}$ should  be taken  as an
upper limit, due to contamination  of the nucleus by stellar light and
obscuration of the H$\beta$ line.

The radio-loudness parameter R$_{\rm  o}$, was calculated as the ratio
of  the 5\,GHz,  mas-scale  radio (except  for  NGC\,7331) to  nuclear
B-band luminosity.

In   comparison,  the  radio-loudness   parameter  R$_{\rm   x}$  was
calculated as the  ratio of the 5\,GHz, mas-scale  radio (except  for  NGC\,7331) to hard X-ray
luminosity, as formulated in Terashima \& Wilson (2003).

The  Eddington  luminosity --  the  maximum  luminosity  output of  an
accreting object in an isotropic homogeneous system -- was calculated as
(in units of erg s$^{-1}$):

\begin{center}

L$_{\rm Edd}$=1.3 $\times$ 10$^{38}$ $\left( \frac{\rm M_{\rm BH}}{\rm
M_{\odot}} \right)$

\end{center}

\subsection{Radio-Loudness} 

Objects  with log  R$_{\rm o}>$1  are commonly  said to  be radio-loud
(e.g.   Kellermann  \etal 1994).   Ho  \&  Peng  (2001) have  recently
challenged the  long-thought notion  that Seyfert galaxies  are mainly
radio-quiet galaxies.   By using  nuclear B-band magnitudes  and radio
powers, they found that as many as 60\% of Seyferts are radio-loud.

An alternative method for parameterizing radio-loudness involves using
the radio to hard X-ray ratio (Terashima \& Wilson 2003).  This method
has the advantage  of suffering less from extinction  than the optical
and allows the identification of  the nucleus by means of high spatial
resolution  observations   (e.g. {\it  Chandra}).   The   equivalent
radio-loud/radio-quiet boundary  is established  at log  R$_{\rm x}$=$-$4.5
(Terashima \& Wilson 2003).

In  Fig.~3 we have  plotted the  logarithm of  the R$_{\rm  x}$ versus
R$_{\rm  o}$  parameters,  where  the  solid lines  denote  the
radio-loud/radio-quiet boundaries.


\setcounter{figure}{2}
\begin{figure}[!h]
\centering
\includegraphics[width=9cm]{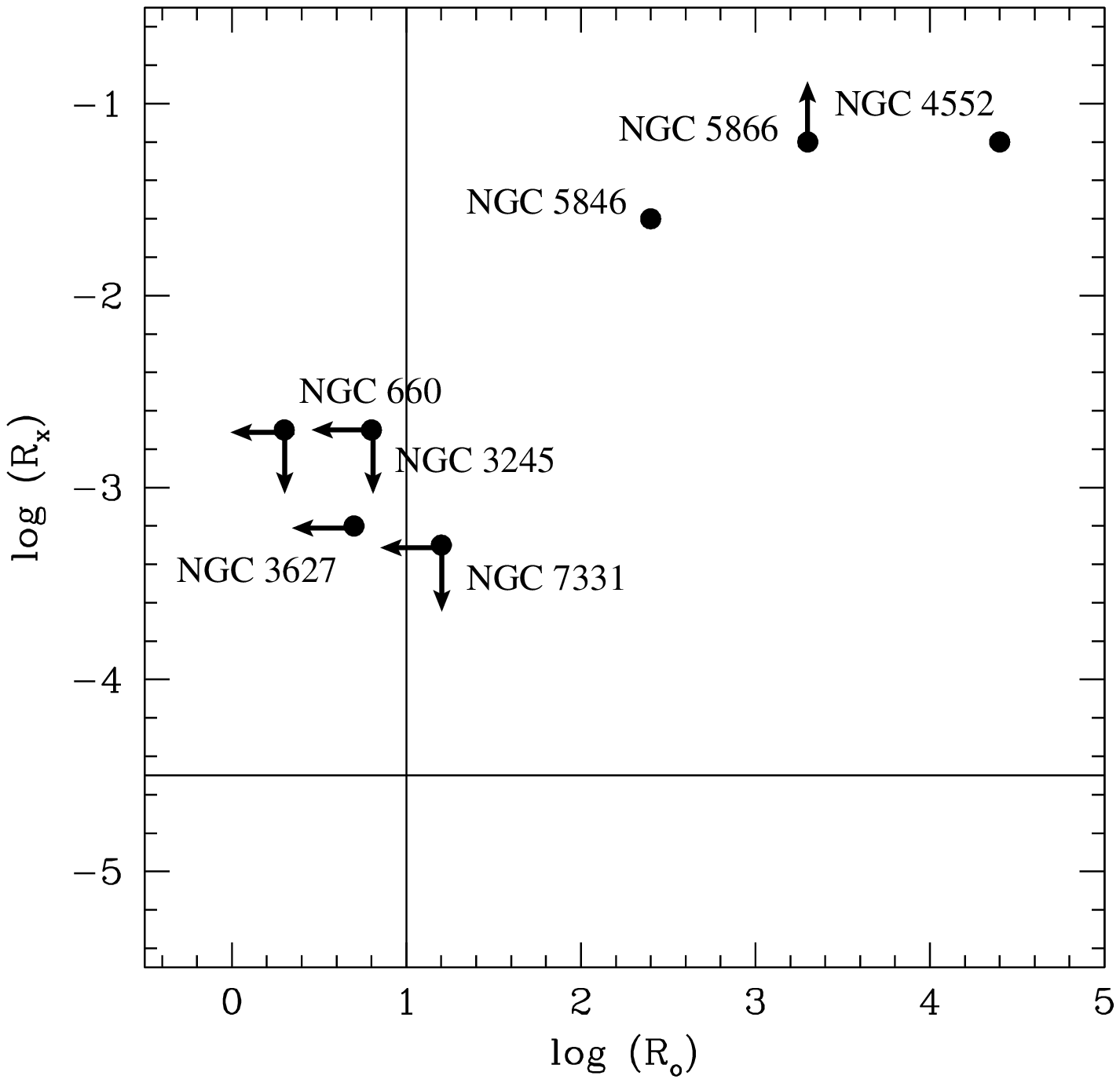}
\caption{The radio to X-ray ratio (R$_{\rm x}$) versus radio to optical
  (R$_{\rm o}$) relation in logarithmic units.  Upper limits in X-ray
and  radio  luminosity are  illustrated  by  arrows.  
 The solid lines define the radio-loud/radio-quiet boundary for the optical
and X-ray diagnostic.}
\end{figure}

NGC\,660, NGC\,3245, NGC\,3627 and NGC\,7331 are found to be optically radio-quiet.  
The  remaining sources  are radio-loud,  according to  the log
R$_{\rm o}>$1 criterion (e.g. Kellermann \etal 1994).  In particular,
all  of  the  mas-scale radio  cores,  as  shown  by the  5\,GHz  VLBA
observations, are radio-loud.  Our results support those of Ho \& Peng
(2001), whereby  radio-loud objects  are also found  in disk-dominated
hosts (see Filho, Barthel \& Ho 2000, 2002).

As can be seen from the  plot as well as Table~4, with the possible exception
of NGC\,660, NGC\,3245 and NGC\,7331,
all  sources are  radio-loud  according to  the  log R$_{\rm  x}>-$4.5
criterion.   In  particular,  all  the detected  mas-scale  cores  are
radio-loud.

The R$_{\rm  o}$ and R$_{\rm x}$  results are consistent,  in that all
mas-scale  radio cores  are  radio-loud. However,  differences in  the
remaining sources  most likely arise from stellar  contribution. It is
to be  noted, for example, the high  H$\alpha$ luminosities associated
with  NGC\,660, NGC\,3245,  and NGC\,3627.  

\subsection{Radio vs. H$\alpha$ and [O~{\sc i}] Luminosity}

It is well  known that classical active galaxies  like Seyferts, radio
galaxies  and  LINERs display  correlations  between  their radio  and
emission-line properties  (Ho \& Ulvestad 2001; Ulvestad  \& Ho 2001a;
Zirbel \&  Baum 1995;  Nagar \etal 2000,  2002). In  particular, Nagar
\etal (2000, 2002)  also show that there is  a correlation between the
radio core luminosity and  [O~{\sc i}]$\lambda$6300 line luminosity in
LINERs and low luminosity Seyferts.

The radio luminosity of the  present sample does not seem to correlate
with  H$\alpha$  or  with  [O~{\sc  i}]$\lambda$6300  luminosity,  but
because  of  the small  sample  and  large  intrinsic scatter  in  the
correlation we are not able to draw any conclusions.  The absence of a
radio and emission-line correlation  for composite sources had already
been  noted in Filho, Barthel \& Ho (2002) using  lower  resolution, VLA  radio
observations.

\subsection{X-Ray vs. H$\alpha$ Luminosity}

By examining the hard X-ray  to H$\alpha$ luminosity ratio, we wish to
investigate  whether the  observed X-ray  emission is  related  to the
continuum source which powers the emission lines.

Such  a correlation between  the X-ray  and H$\alpha$  luminosities is
observed in  classical AGN  such as Seyferts  and quasars  (Ward \etal
1988) and more recently in LINER  1s (Terashima, Ho \& Ptak 2000; Ho
\etal  2001), indicating  that  the  AGN is  capable  of powering  the
optical emission lines. Typical ratios for LLAGN are in the range:

\begin{center}

$\rm {log \left(\frac{L_{\rm X}}{L_{\rm H\alpha}}\right) \sim 1 - 2 }$

\end{center}

In Fig.~4 we plot the relation for our sample sources, where the X-ray
luminosity is taken from the  hard X-ray band.  The shaded area defines ratios
that are typical for AGN-powered sources.


\setcounter{figure}{3}
\begin{figure}[ht!]
\centering
\includegraphics[width=9.0cm]{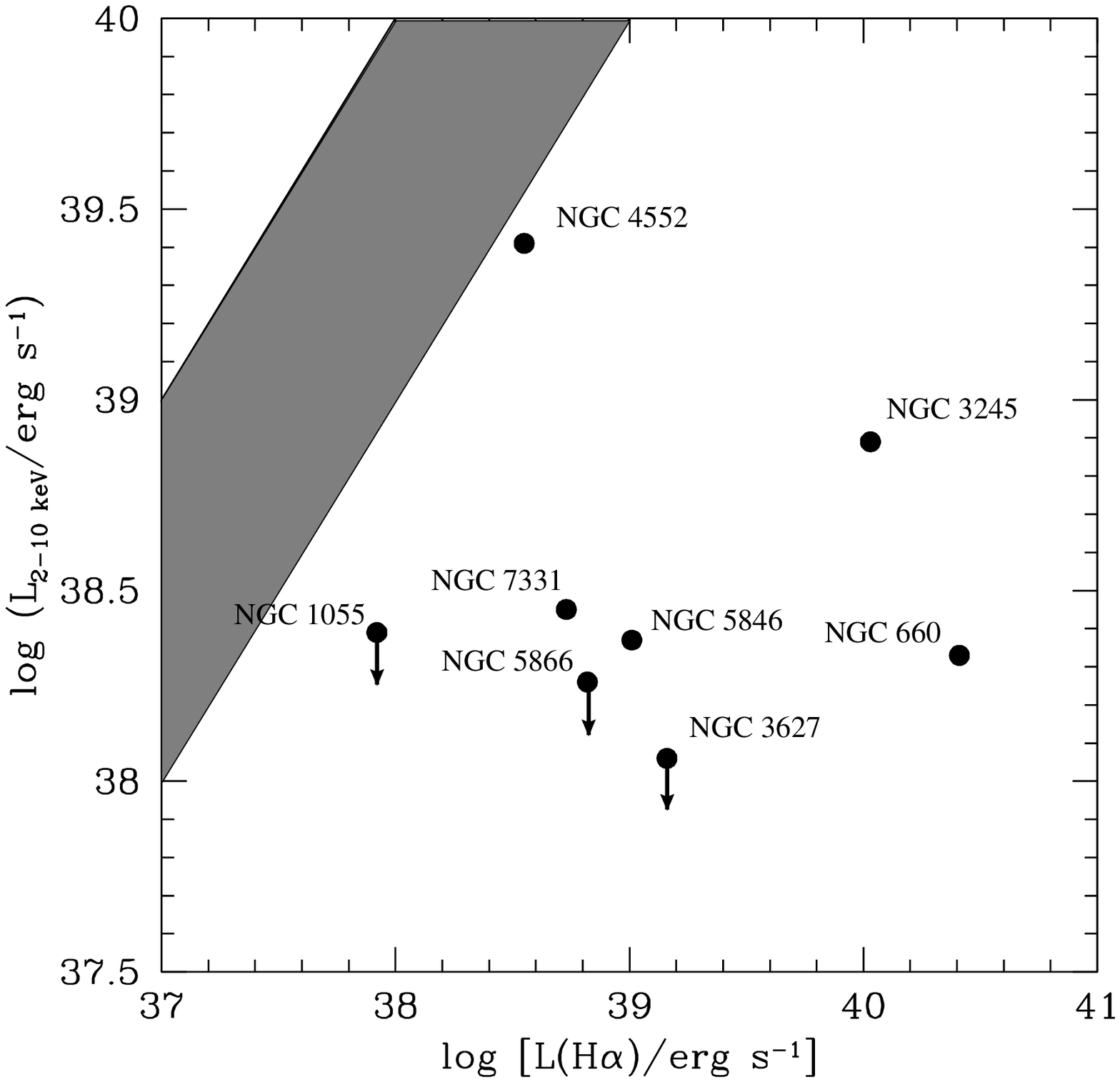}
\caption{The  relation between hard X-ray core and H$\alpha$ emission
  line luminosity.
Upper limits in X-ray luminosity are illustrated by arrows. 
The shaded area corresponds to typical
values for LLAGN --  ${\rm log \left(\frac{L_X}{L_{H\alpha}}\right) \sim 1 - 2}$.}
\end{figure}

All  of the sources  fall short  of the  empirical X-ray  to H$\alpha$
ratio observed  in classical AGN,  indicating that the  {\it observed}
X-ray luminosities are too low to generate the H$\alpha$ lines.  There
are three possible scenarios --  either the emission lines are powered
by a source  other than the AGN, the ionizing spectrum is different than that inferred,
or the X-ray  emission  is heavily
obscured  at energies above  2~keV.  The  first possibility has
already  been investigated  for  example  in the  work  of Maoz  \etal
(1998), whereby  circumnuclear stellar populations  may provide enough
ionizing photons  to explain the observed  optical emission-line flux.
However, as investigated by Ho,  Filippenko \&
Sargent (2003; see also Ho  \etal 2001), there is no evidence
for the  presence of young stellar clusters. Alternatively, it may
be that the underlying ionizing spectrum is more complicated then
assumed -- most of the ionizing photons may occur in the extreme
ultraviolet (EUV) and soft X-ray region, not in the hard X-rays. 
Powering of the  emission lines
from a central AGN could still occur if the hard X-ray flux were being
heavily  absorbed (Terashima,  Ho \&  Ptak 2000;  Terashima  \& Wilson
2003). If {\it intrinsic} X-ray luminosities  are 1 -- 2 orders of magnitude
higher, then the  X-ray to H$\alpha$ ratio would  be typical for LLAGN
(Terashima  \& Wilson 2003).   Although scattering  is still  a viable
explanation, observations at energies above 10~keV are needed in order
to observe any Compton-thick component.

\subsection{Radio vs. X-ray Luminosity}

In Fig.~5  we plot  the mas-scale radio  core luminosity  versus the
hard X-ray core luminosity for the sample sources.


\setcounter{figure}{4}
\begin{figure}[h!] 
\centering
\includegraphics[width=9.0cm]{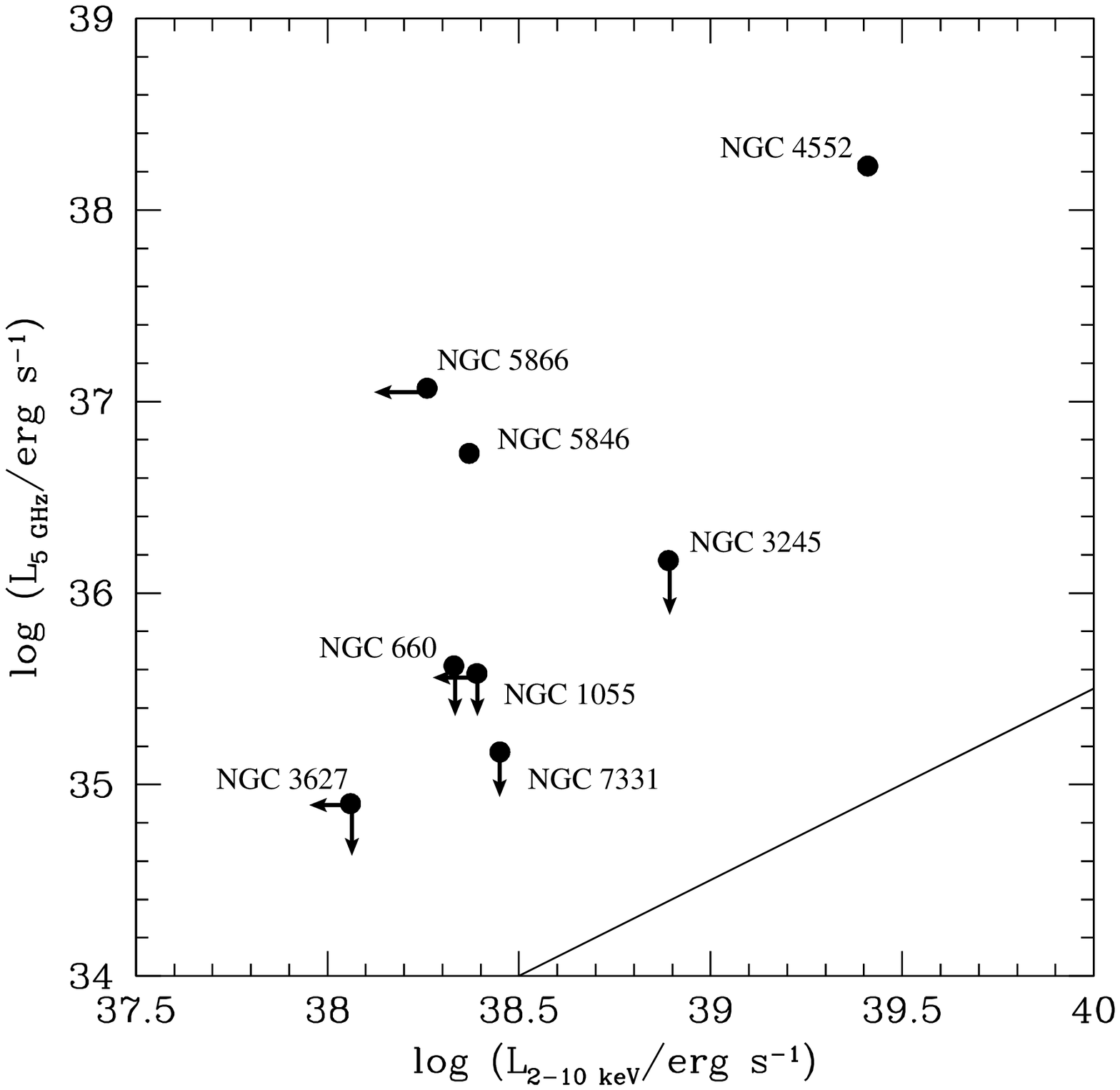} 
\caption{The  relation between the  mas-scale radio core and hard X-ray core luminosity.
Upper  limits  in  radio  and  X-ray  luminosity  are  illustrated  by
arrows.  
The    solid    line    defines    the
radio-loud/radio-quiet boundary for the X-ray diagnostic.}
\end{figure}

It  can be  seen from  the  plot that  the detected  mas-scale radio cores  have
L$_{\rm  R}>10^{36.5}$ erg  s$^{-1}$, with  two sources  (NGC\,4552 and NGC\,5846) being firm
{\it Chandra}  detections (Table~3).  
Three  of  the sources with no  mas-scale radio detection,  however,  show hard  X-ray  {\it  Chandra}
detections  -- NGC\,660, NGC\,3245  and NGC\,7331. 
The {\it  Chandra} observations of NGC\,1055 and NGC\,3627 (see also Ho
\etal 2001) failed to detect any hard X-ray nuclear source,
most likely due to the low integration
time for the X-ray observations ($\tau$$<$2~ks; Table 3).  Combining
both the radio and X-ray data yields a $\sim$50\% LLAGN detection rate among the sample sources
or equivalently a 12\% detection rate for the entire composite LINER/H~{\sc ii} galaxy sample.

NGC\,7331
has  a very  low luminosity  VLA radio core  (Cowan, Romanishin  \& Branch
1994), 10 times that of Sgr A$^*$, and also the lowest hard X-ray core
luminosity of all the sample  sources, 10$^4$ times that of Sgr A$^*$.   
Similarities   in  the   arcsec-scale  radio
properties -- radio morphology and  low radio flux density -- of NGC\,7331 and
the  cores  that  were  not  detected  on  mas-scales,  indicate  that
NGC\,7331 may  actually be  a prototype. These sources may all harbour  
submillijansky radio and low luminosity hard X-ray AGN
cores, difficult to detect unless deep observations are obtained. Nevertheless, the results 
demonstrate the combined power of high resolution 
X-ray and radio imaging in LLAGN detection. We will return to this issue in the
next section. 

\section{ADAFs or Jets?}

We wish to use our observations to test models of radio and
X-ray emission in low luminosity sources, namely advection-dominated accretion 
flows (ADAFs) and jet
models. These models  have been extensively applied to several
other low luminosity sources, of  which Sgr A$^{*}$ (Yuan, Markoff \&
Falcke 2002; Yuan, Quataert \& Narayan 2003)  and NGC\,4258 (Yuan \etal 2002)  are just
two examples.

ADAFs (see reviews by Narayan, Mahadevan \& Quataert 1998 and Quataert
2001) occur    in   the   low    accretion   rate    regime   (m$_{\rm
acc}$(crit)$<$10$^{-1.6}$,   where   m$_{\rm   acc}$   is   given   in
dimensionless   units   of  Eddington   accretion   rate).  They   are
characterized by a low radiative efficiency, due to the advection of a
significant amount of the energy into the black hole.  ADAF theory has
developed quickly in  recent years with the inclusion  of outflows and
the possible existence of  nonthermal electrons (see Yuan, Quataert \&
Narayan  2003 for  a  review).  However,  the  analysis and  therefore
conclusions of this paper are based on the canonical ADAF model.

ADAFs may be able to account
for  correlations  observed  between  the radio  and  X-ray  emission.
Particularly, this type of  low radiative efficiency flow is supported
by the  highly sub-Eddington luminosities observed in  our sources
($\frac{\rm  L_{X}}{\rm  L_{Edd}}<10^{-5}$;  Table~4).  Furthermore,
ADAFs are naturally  bright in radio due to  radio synchrotron cooling
and dim in the UV/optical regime due to the absence of a `blue bump',
the signature of a standard, optically thick but geometrically thin  accretion disk
(Shakura \&  Sunyaev 1973). Both
of these ADAF characteristics conspire to  produce `radio-loud' cores. 

In  ADAFs  the  hard  X-ray emission  arises  from bremsstrahlung  and
Compton scattering of  synchrotron photons subject to self-absorption.
As  a  result,the  ADAF model  predicts  the
correlation (Yi \& Boughn 1998, 1999):

\begin{center}

L$_{\rm R} \propto {\rm M_{BH}}\,{\rm L_{X}^{0.1}}$

\end{center}

In  Fig.~6  we  plot  the
mas-scale radio  core to hard X-ray ratio versus the 
hard X-ray  core  luminosity.   In parentheses  are the logarithms of the
black  hole mass plus error estimations,  which include  the error  in
velocity dispersion  plus a 10\% error associated  with the conversion
from  $\sigma_e$ to  $\sigma_c$. The  lines  refer to  the ADAF  model
prediction for  different black  hole masses, appropriately  scaled to
5\,GHz.


\setcounter{figure}{5}
\begin{figure}[!ht]
\centering 
\includegraphics[width=9.0cm]{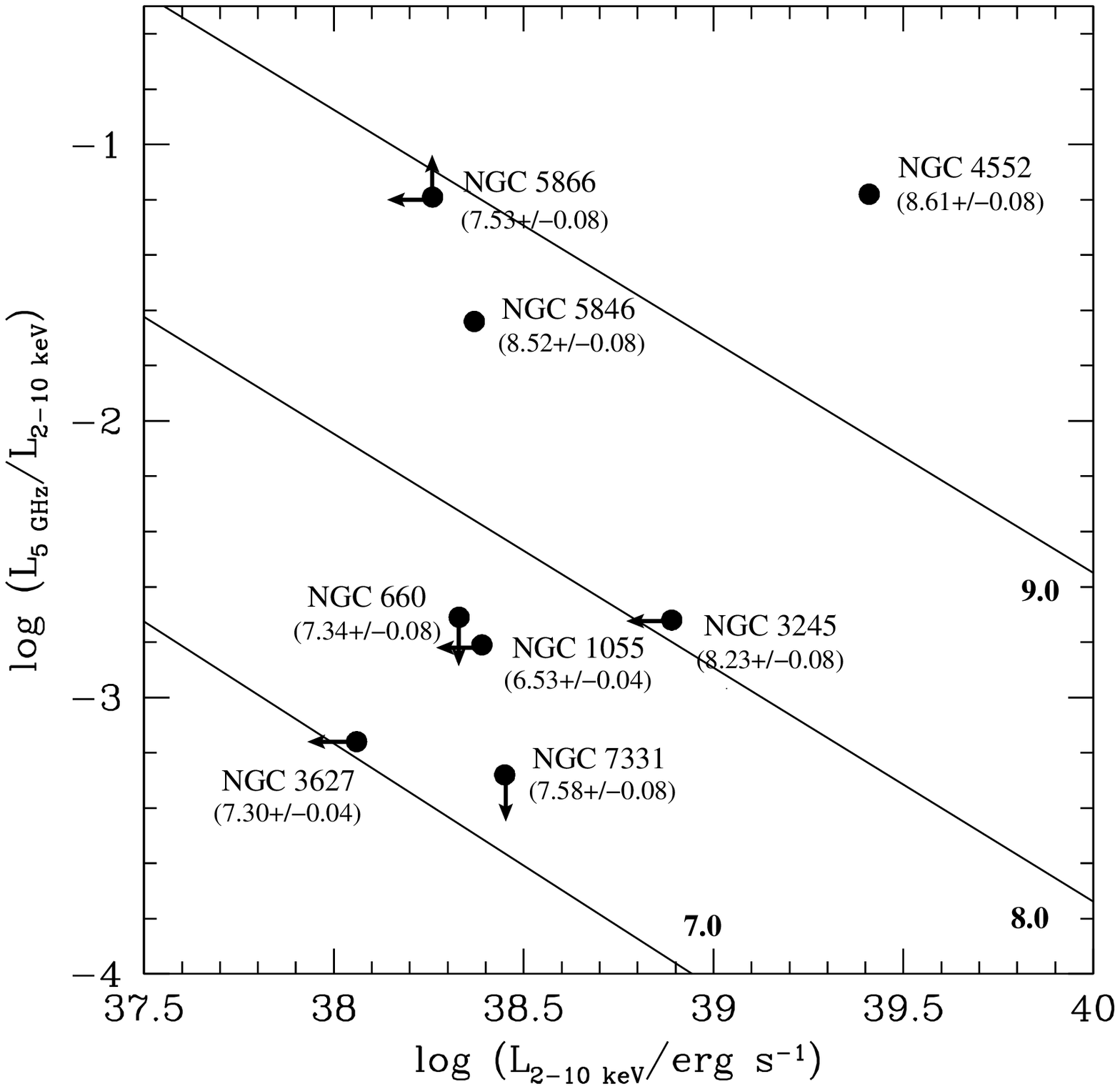} 
\caption{The relation between the mas-scale radio
core/hard  X-ray  core   luminosity  ratio  (L$_{\rm  5\,GHz}$/L$_{\rm
2-10 keV}$) and the hard X-ray  core luminosity.  Upper limits in radio
and  X-ray  luminosity are  illustrated  by  arrows.  
Numbers  in parentheses  refer to the  logarithm of  the black
hole masses and the respective error, considering an error in log
$\sigma_c$=0.06, 0.02 and 0.01 plus a 
10\% error in the conversion of $\sigma_e$ to $\sigma_c$.  The solid lines refer to the
ADAF model prediction -- ${\rm L_{R} \propto M_{BH}\, L_{X}^{0.1}}$
--  for several black  hole masses  -- ${\rm  log \,  M_{BH}=7.0, 8.0,
9.0}$ -- in logarithmic units of solar mass.}
\end{figure}

If the  sources with  upper limits to  their mas-scale  radio emission
contain accreting black  holes as is the case for  NGC\,7331 (due to a
hard  X-ray detection),  then it  is possible  to explain  their radio
emission using the ADAF model,  as originating in underfed black holes
(m$_{\rm  acc}<10^{-3}$).  However, for  the mas-scale  detected radio
cores, the ADAF  model overpredicts the black hole  masses by a factor
10 or so (Table~4) even when the errors associated with the black hole
mass  determination are  considered (Fig.~6).  This suggests  that the
ADAF model is not entirely consistent with the observed radio to X-ray
ratio unless the hard X-rays are heavily obscured (see Section~5.3).

In Fig.~7  we plot  the mas-scale radio  core luminosity  versus the
black hole mass. The solid lines refer to the ADAF prediction of:

\begin{center}

L$_{\rm R} \propto {\rm m_{acc}}^{1.2}\, {\rm M_{BH}^{1.6}}$

\end{center}

\noindent where the relation has been appropriately scaled to 5\,GHz.
Typical errors in black hole mass are given. 



\setcounter{figure}{6}
\begin{figure}[!h]
\centering
\includegraphics[width=9.0cm]{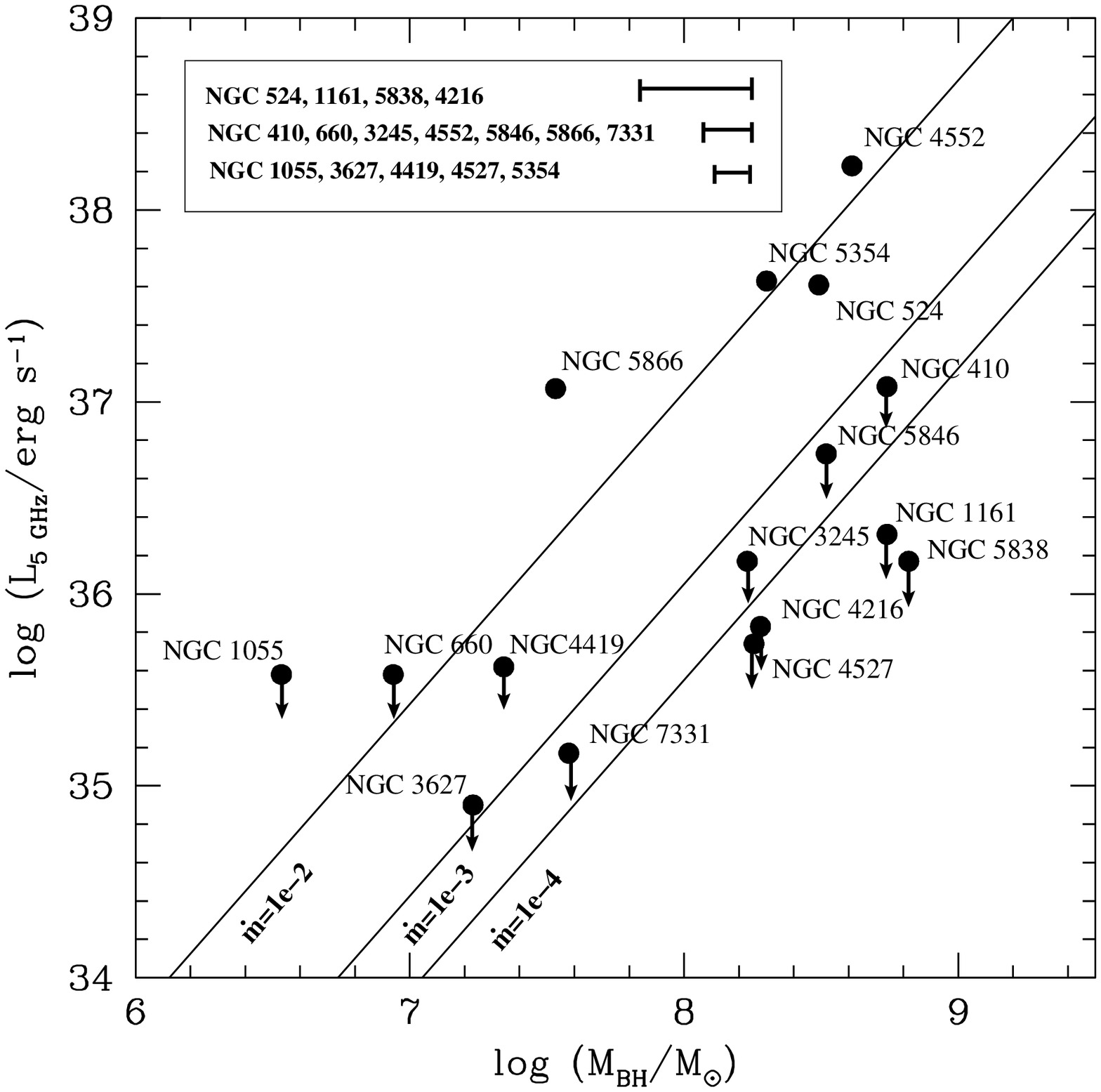} 
\caption{The  relation between the mas-scale
radio core luminosity and the  black hole mass.  Upper limits in radio
luminosity are  illustrated by  arrows. We give typical error bars for the black hole mass,
considering an error in log $\sigma_c$=0.06, 0.02 and 0.01 plus a 
10\% error in the conversion of $\sigma_e$ to $\sigma_c$. The solid  lines refer  to the
ADAF     model    prediction     --    ${\rm     L_{R}    \propto
m_{acc}^{1.2}\,M_{BH}^{1.6}}$ -- for different mass accretion rates --
${\rm  m_{acc}=10^{-2}, 10^{-3}, 10^{-4}}$  -- given  in dimensionless
units of Eddington accretion rate. }
\end{figure}

The  results show that  composite sources  with mas-scale  radio cores
have  a radio  luminosity  to black  hole  mass ratio  similar to  low
luminosity  Seyferts and  LINERs (log  (L$_{\rm R}$/M$_{\rm BH}$)
$\sim$ 1.14;  Nagar \etal  2002).  On the  other hand, if  the sources
with no mas-scale radio detection harbour an AGN, then on average they
should have  lower radio to  black hole ratios  than the LLAGN  in the
Nagar \etal (2002) sample.

Fig.~7 also  shows the ADAF  model prediction for  several accretion
rates.    Sources
with  upper limits to  their mas-scale  radio emission  are consistent
with low mass accretion rates (m$_{\rm acc}<10^{-2}$).  However, the
detected  mas-scale  cores  fall  very  close to  the  ADAF
permitted accretion limit of m$_{\rm acc}({\rm crit})=10^{-1.6}$.  
These results suggest  that the  mas-scale radio  cores show
more radio emission than can perhaps be accounted for from an ADAF.

Plotted in Fig.~8 is the  relation between the hard X-ray luminosity
and  the black  hole  mass. The  solid  lines refer  to the  predicted
correlation in an ADAF model for various mass accretion rates:

\begin{center}

L$_{\rm X}$ $\propto$  M$_{\rm BH}$ m$_{\rm acc}^{\rm x}$

\end{center}

\noindent where x=2 for thermal emission. Again, typical errors in black hole mass
determination are given in the plot.



\setcounter{figure}{7}
\begin{figure}[!ht]
\centering
\includegraphics[width=9.0cm]{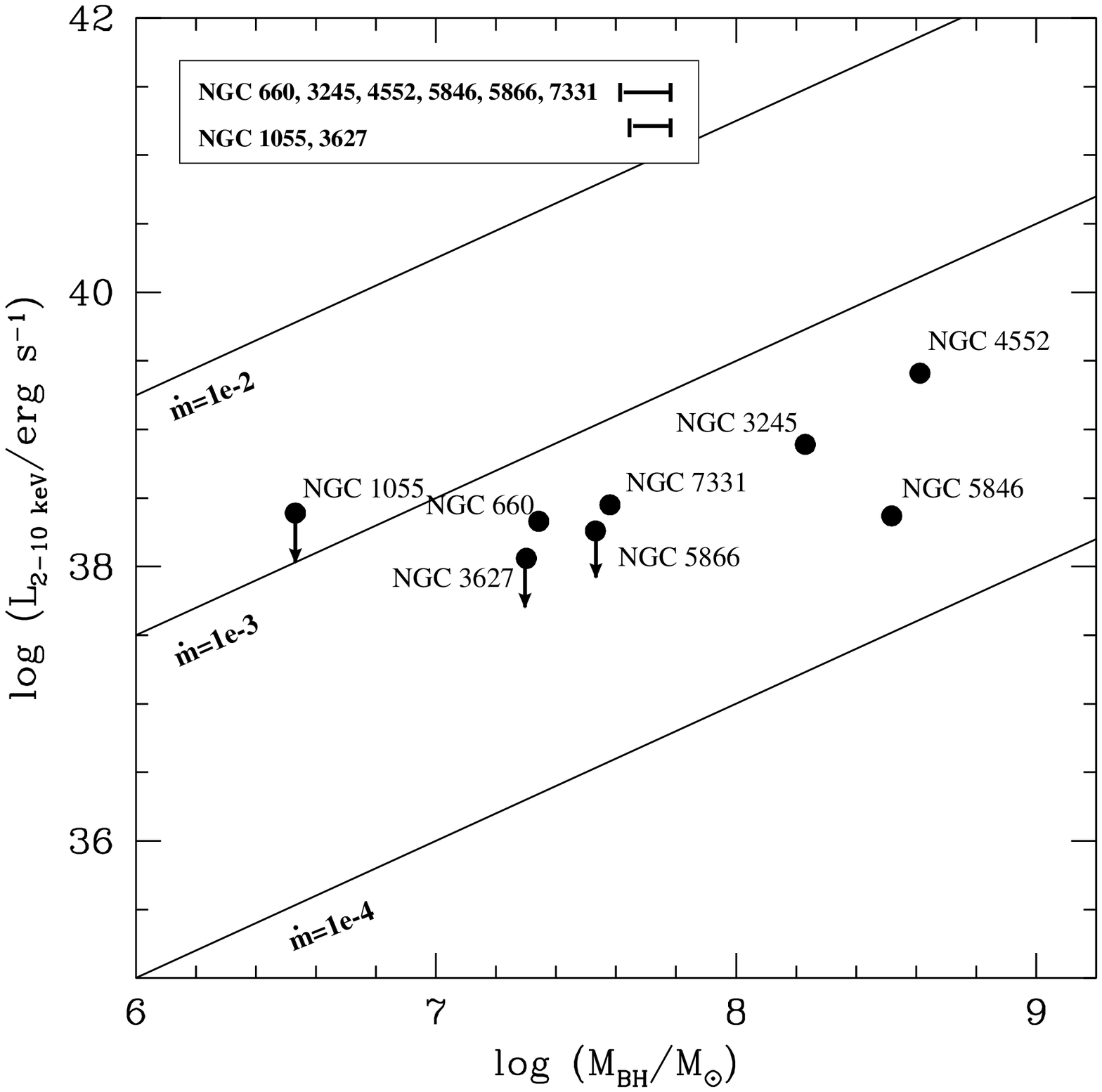} 
\caption{The relation between the hard X-ray core
luminosity and black hole mass.   Upper limits in X-ray luminosity are
illustrated by arrows.  We give  typical error bars for the black hole
mass, considering an error  in log $\sigma_c$=0.06, 0.02 and 0.01 plus
a 10\% error in the  conversion of $\sigma_e$ to $\sigma_c$. The solid
lines  refer to  the  ADAF  model prediction  --  ${\rm L_{X}  \propto
M_{BH} m_{acc}^{2}}$   -- for  various  accretion rates  -- ${\rm  m_{acc}=10^{-2},
10^{-3},  10^{-4}}$  --  given  in dimensionless  units  of  Eddington
accretion rate. }
\end{figure}

The X-ray  data are consistent with  the ADAF model  for low accretion
rates  (m$_{\rm  acc}<10^{-3}$),  both   for  the  sources  with
mas-scale upper limits {\it and}  for those  with mas-scale  cores. This
could suggest  the same underlying  mechanism for X-ray  production in
all sample sources -- an ADAF.  In fact, in two of the mas-scale radio
cores  where  an  X-ray  spectral  fit  was  possible  (NGC\,4552  and
NGC\,7331; see  Table~3), the X-ray photon index  is consistent with
the predicted ADAF value of $\sim$1.4.

In summary, if we take NGC\,7331 as a prototypical submillijansky AGN
core, then scaling arguments suggest that an ADAF could simultaneously
explain both the radio and X-ray emission in the sources with no
mas-scale radio detection.  In fact, these sources may in fact be
`bare' ADAFs, where the radio and X-ray contribution is coming mainly
from the accretion flow.  The absence of a jet or of an energetically
significant contribution of a jet to the radio emission is not,
however, clear. It may be that, like in M\,81, the jet is present but
very compact and difficult to detect (Bietenholz, Bartel \& Rupen 2000).

On the other hand, thermal ADAF models predict inverted radio spectra
($\alpha \sim - $ 0.4; Narayan, Mahadevan \& Quateart 1998; see also Yuan, Quataert \&
Narayan 2003) while the mas-scale sources possess flat spectrum radio
cores (Table~2). Modified ADAF models that include convection (Ball,
Narayan \& Quataert 2001) or outflows (Quataert \& Narayan 1999) do not provide
satisfactory solutions, since the predicted spectral slopes are roughly the same 
as in the canonical  ADAF models (Quataert \& Narayan
1999). \footnote {The inclusion of power-law electrons has been recently proven to 
produce a flatter radio
spectra (see Yuan, Quataert \& Narayan 2003).}  
Therefore, while the X-ray emission in mas-scale radio cores
may be ADAF-powered, the radio emission may require another emission
mechanism.

The most likely scenario is that the radio emission in the mas-scale
detected radio cores is a result of an ADAF-type accretion (or
variants thereof) plus a compact jet or outflow (Quataert \etal 1999; 
Yuan 2000; Falcke \etal 2001; Nagar, Wilson \& Falcke 2001;
Ulvestad \& Ho 2001b; Nagar \etal 2002).  The consideration of a jet
is not ad-hoc since we detect extended emission in 4 out of the 5
mas-scale radio cores.  Furthermore, the addition of a self-absorbed
jet component to an ADAF core may `flatten' out the spectral index to
the observed values. In fact, flat spectrum radio emission in
`classical' AGN are commonly attributed to the presence of jets.           

In a recent paper, Falcke, K\"ording \& Markoff (2003; see also
Merloni \etal 2003) investigated an accretion rate-dependent scheme
for black hole-powered sources, whereby sub-Eddington systems are
predicted to be dominated by emission from a relativistic jet. Such an
idea was suggested already in Falcke \& Biermann (1995), and a jet
model was shown to analytically predict the radio/X-ray correlation
seen in the low/hard state of the black hole candidate X-ray binary
(XRB) GX\,339--4 (Markoff \etal 2003).  This correlation has now been
seen in many other low/hard state XRBs, and is likely universal
(Gallo, Fender \& Pooley 2003).
Falcke, K\"ording \& Markoff (2003; see also Merloni \etal 2003) show 
that this correlation can be
scaled with black hole mass to extend to certain classes of AGN as well.  This
leads to a radio/X-ray/black hole mass fundamental plane which unifies
diverse sources such as FR I radio galaxies, LINERs and XRBs as
jet-dominated (see Fig. 1 in Falcke, K\"ording \& Markoff 2003).

This scaling relation is given by (eq. 7; Falcke, K\"ording \& Markoff
2003):

\begin{center}

L$^{\rm cor}_{\rm X}$ $\propto$ L$_{\rm X'} \left(\frac{\nu_{\rm
  X'}}{\nu_{\rm X}}\right)^{-\alpha_{\rm X}}$  M$_{\rm BH}^{-m\alpha_{\rm R}+\alpha_{\rm X}}$

\end{center}

\noindent where $\alpha_{\rm R}$ and $\alpha_{\rm X}$ are the radio
and X-ray spectral indices and $m$ is a function of both $\alpha_{\rm R}$ and
$\alpha_{\rm X}$.  We here apply the black hole mass scaling
relation in order to investigate the presence of an energetically
dominant jet in our sample sources. We have extrapolated our 2--10 keV
luminosities (Table~3) to 3--9 keV assuming  an average $\alpha_{\rm X}$=0.6 ($\equiv \Gamma-$1) and $\alpha_{\rm R}$=-0.15 (Falcke, K\"ording \& Markoff
2003). The assumption  of a spectral index of $\Gamma$=1.6 ($\alpha_{\rm X}$=0.6) is consistent within 10\% of values obtained using our actual tabulated spectral indices (Table~3) and therefore does not alter the general result.

In Fig. 9 we have plotted the scaled correlation normalized to the 6
M$_{\odot}$ black hole mass of GX\,339--4. Typical errors in the
`black hole mass-corrected' X-ray luminosity are conservatively estimated 
to be of the order
of 0.08 in logarithmic units of luminosity.


\setcounter{figure}{8} 
\begin{figure}[!h] 
\centering
\includegraphics[width=9.0cm]{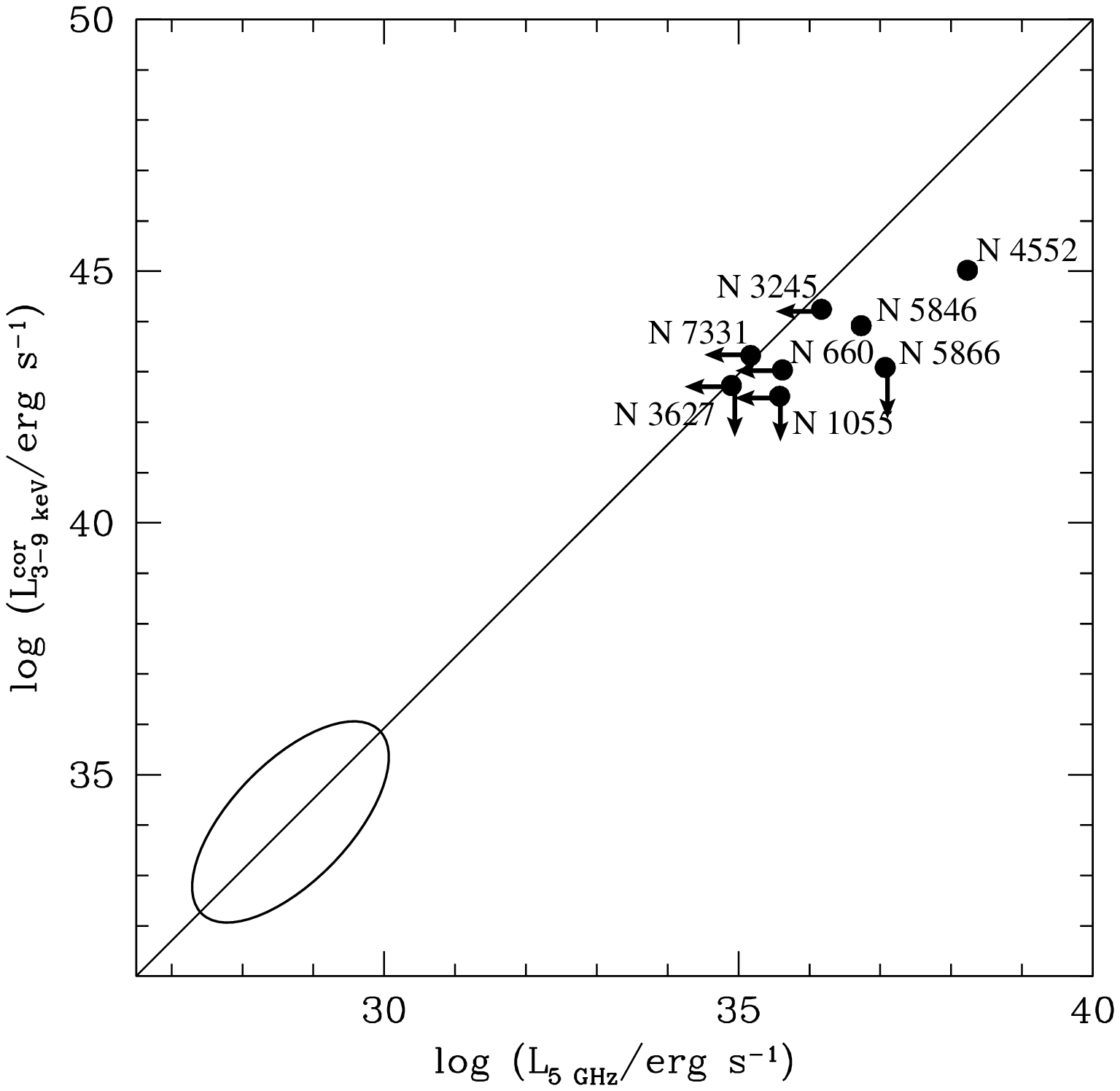}
\caption{The relation between radio and `black hole mass-corrected'
X-ray luminosity, scaled to the 6 M$_{\odot}$ black hole in XRB
GX\,339--4. The solid line is the predicted radio/X-ray correlation
from the jet model and the ellipse details the broadband observations
of GX\,339--4 (Corbel et al. 2003; Markoff et al. 2003).}  
\end{figure}      

In addition to the scenario discussed above, a jet model alone seems
to interpret the present data  as well.
It is noteworthy that the mas-scale detected sources fall very close
to the jet model predicted relation.  
The deviation of the sources NGC\,4552, NGC\,5846 and
NGC\,5866 is in the direction of {\it excess} radio emission, which is
entirely consistent with the jet model. The scaling relation is valid
for the {\it compact} inner jet core while these sources are known to possess
jet emission on mas-scales; hence, the radio flux density likely
includes some extended jet emission.     


\section{Conclusions}

We have  embarked on a high resolution X-ray and radio imaging  study of a  sample of
composite  LINER/H~{\sc   ii}  galaxies  known   to  exhibit  AGN-like
properties.   Five of  the 16  sample sources possess milliarcsecond radio
cores, 
four  of  which also show extended radio emission. Two of the sources with no milliarcsecond
radio detection were found to possess hard X-ray nuclear cores. The 50\% detection rate of
radio and/or hard X-ray nuclei support the view that these AGN candidate composite H~{\sc   ii}  galaxies 
host weak AGN central engines. When considering the entire composite  LINER/H~{\sc   ii} sample, these results yield an overall 12\%  detection rate.
Modeling of the radio  and X-ray emission suggests that the mas-scale radio detected
sources exhibit an energetically important contribution from a
jet, associated with a sub-Eddingtion accreting black hole. 
The sources  that were not detected on  mas-scales, however, appear
consistent with the radio and X-ray emission coming mainly from the low
radiative efficiency accretion flow (ADAF). 
Nonetheless,  even  in the  cases  where the  presence  of  an AGN  is
irrefutable, this AGN cannot be responsible solely for powering the observed
emission lines unless the X-ray flux is heavily obscured.


\begin{acknowledgements}

M.~E.~F. acknowledges support from the Funda\c c\~ao para a Ci\^encia e
Tecnologia, Minist\'erio  da Ci\^encia e Ensino Superior, Portugal through
the grant PRAXIS XXI/BD/15830/98 and SFRH/BDP/11627/2002. We would like to thank Heino Falcke
for the extensive discussions on ADAF and jet models and 
Jim Ulvestad for providing us with calibrated VLA data for NGC\,660.
We wish also to thank the anonymous referee for his helpful comments.

This research  has made extensive use of  NED (NASA/IPAC Extragalactic
Database),  which  is  operated  by  the  Jet  Propulsion  Laboratory,
California Institute  of Tecnology,  under contract with  the National
Aeronautics and Space  Administration, LEDA (Lyon/Meudon Extragalactic
Database), HYPERCAT and  the {\it Chandra} data archive.   The VLA and
the VLBA  are facilities of  the National Radio  Astronomy Observatory
(NRAO) which is a facility of the National Science Foundation operated
under cooperative agreement by Associated Universities, Inc.

\end{acknowledgements}







\clearpage


\begin{table*}

\scriptsize

\begin{center}

\sbox{\rotbox}{

\setcounter{table}{0}

\begin{minipage}{187mm}

\caption{{\bf Map parameters (published or new) of the sample sources.} 
Col. 1: Source name.
Col. 2: Observing frequency.
Col. 3: Integration time.
Col. 4: $rms$ noise level of the image.
Col. 5: Restoring  beam or resolution.
Col. 6: Position angle  of the  beam.
Col. 7: Peak radio flux density.
Col. 8: Integrated radio flux density.
Col. 9 and 10: Radio position.
Col. 11: Deconvolved source size.
Col. 12: Position angle of the source. Value in parenthesis is an uncertain value.
Col. 13: Note. Letter is the observing epoch and number is the
reference (published or new) from which the radio data were taken.
}
\smallskip

\begin{tabular}{l c c c c cc  cc cc c c c c c}

\hline
\hline
 
 & $\nu$ & $\tau$ & $rms$ & Beamsize &  \multicolumn{2}{c}{P.A.} & 
 \multicolumn{2}{c}{F$_{\rm peak}$} &  \multicolumn{2}{c}{F$_{\rm int}$} & RA(J2000) & Dec(J2000) & Size & P.A. &   \\
Galaxy & (GHz) &  (min) & ($\frac{\rm mJy}{\rm beam}$) & 
(arcsec$^2$) &  \multicolumn{2}{c}{(deg)} &
\multicolumn{2}{c}{($\frac{\rm mJy}{\rm beam}$)} & \multicolumn{2}{c}{(mJy)} & (\hrs \ramin \rasec) & 
(\degree \arcmin \arcsec~) & (arcsec$^2$) & (deg) & Note \\
(1) & (2) & (3) & (4) & (5) &  \multicolumn{2}{c}{(6)} & 
\multicolumn{2}{c}{(7)} & \multicolumn{2}{c}{(8)} & (9) & (10) & (11) &  (12) & (13) \\

\hline

VLA Obs. & & & & & \multicolumn{2}{c}{} &\multicolumn{2}{c}{} &
\multicolumn{2}{c}{} & & & & & \\

\hline

NGC\,660  & 8.4 & 270 & 0.01 & 0.21$\times$0.21 & \multicolumn{2}{r}{$-$39.90} & \multicolumn{2}{r}{0.5} &
\multicolumn{2}{r}{2.3} & 01 43 02.32 & +13 38 44.9 & 0.46$\times$0.36 & 174.30 & 1,a \\  
       
          & 15 & 16 & 0.20 & 0.15 & \multicolumn{2}{r}{\ldots} & \multicolumn{2}{r}{$<$0.9} & \multicolumn{2}{r}{$<$0.9} & \ldots & \ldots & \ldots & \ldots & 2 \\

NGC\,1055 & 15 & 16 & 0.20 & 0.15 & \multicolumn{2}{r}{\ldots} & \multicolumn{2}{r}{$<$1.8} & \multicolumn{2}{r}{$<$1.8} & \ldots & \ldots & \ldots &  \ldots & 2 \\ 

NGC\,3627 & 15 & 16 & 0.20 & 0.15 & \multicolumn{2}{r}{\ldots} &  \multicolumn{2}{r}{1.1} &  \multicolumn{2}{r}{2.9} & 11 20 15.00 & +12 59 29.6 & \ldots & \ldots & 2 \\

         

NGC\,4216     & 15 & 16 & 0.20 & 0.15 & \multicolumn{2}{r}{\ldots} &  \multicolumn{2}{r}{1.2}    &  \multicolumn{2}{r}{1.3}     & 12 15 54.37 & +13 08 58.0 & \ldots & \ldots & 2  \\


NGC\,4419     & 15 & 16 & 0.20 & 0.15 & \multicolumn{2}{r}{\ldots} &  \multicolumn{2}{r}{2.7}    &  \multicolumn{2}{r}{3.6}    & 12 26 56.45 & +15 02 50.7 & \ldots & \ldots & 2 \\ 


NGC\,4527 & 15 & 16 & 0.20 & 0.15 & \multicolumn{2}{r}{\ldots} & \multicolumn{2}{r}{$<$1.1} & \multicolumn{2}{r}{$<$1.1} & \ldots & \ldots & \ldots & \ldots & 3  \\

NGC\,4552 & 8.4 & 10 & 0.15 & 0.25$\times$0.21 & \multicolumn{2}{r}{$-$27.79} & \multicolumn{2}{r}{102.1} & \multicolumn{2}{r}{102.2} & 12 35 39.80 & +12 33 22.7 & \ldots & \ldots & 4,b \\

            & 15 & 16 & 0.20 & 0.15 & \multicolumn{2}{r}{\ldots} &  \multicolumn{2}{r}{58.1} &  \multicolumn{2}{r}{58.6} & 12 35 39.80 & +12 33 22.7 & \ldots  & \ldots & 2 \\

NGC\,5354 & 8.4 & 10 & 0.12 & 0.26$\times$0.24 &  \multicolumn{2}{r}{73.41} &  \multicolumn{2}{r}{9.7}
&  \multicolumn{2}{r}{10.1}
& 13 53 26.70 & +40 18 10.0 & \ldots &  15.73 & 4,b \\

NGC\,5838 & 8.4 & 10 & 0.07 & 0.31$\times$0.23 &  \multicolumn{2}{r}{36.77} & \multicolumn{2}{r}{1.6} &
 \multicolumn{2}{r}{1.6} & 15 05 26.24 & +02 05 57.4 & \ldots & \ldots & 4,b \\

NGC\,5846 & 8.4 & 10 & 0.10 & 0.30$\times$0.23 &  \multicolumn{2}{r}{35.29} &  \multicolumn{2}{r}{6.3}
&  \multicolumn{2}{r}{6.4} & 15 06 29.28 & +01 36 20.4 & \ldots & \ldots & 4,b \\  

NGC\,5866 & 15 & 16 & 0.20 & 0.15 & \multicolumn{2}{r}{\ldots} &  \multicolumn{2}{r}{7.1} &  \multicolumn{2}{r}{7.5} & 15 06 29.49 & +55 45 47.5 & \ldots & \ldots & 3  \\

NGC\,7331 & 1.5 & \ldots & 0.03 & 1.8$\times$1.4 & \multicolumn{2}{r}{\ldots} &  \multicolumn{2}{r}{0.2}
& \multicolumn{2}{r}{0.23} & 22 37 04.06 & +34 24 56.9 & \ldots & \ldots & 5 \\ 

          & 5 & \ldots & 0.02 & 1.8$\times$1.4 & \multicolumn{2}{r}{\ldots} &  \multicolumn{2}{r}{0.1} &
           \multicolumn{2}{r}{0.12} & 22 37 04.06  & +34 24 56.9 & \ldots & \ldots & 5 \\ 

          & 15 & 16 & 0.20 & 0.15 & \multicolumn{2}{r}{\ldots} & \multicolumn{2}{r}{$<$1.1} & \multicolumn{2}{r}{$<$1.1} & \ldots
          & \ldots & \ldots & \ldots & 2 \\

\hline
\hline
 
 & $\nu$ & $\tau$ & $rms$ & Beamsize & \multicolumn{2}{c}{P.A.} & 
\multicolumn{2}{c}{F$_{\rm peak}$} & \multicolumn{2}{c}{F$_{\rm int}$} & RA(J2000) & Dec(J2000) & Size & P.A. &   \\
Galaxy & (GHz) &  (min) & ($\frac{\rm mJy}{\rm beam}$) & 
(mas$^2$) & \multicolumn{2}{c}{(deg)} &
\multicolumn{2}{c}{($\frac{\rm mJy}{\rm beam}$)} & \multicolumn{2}{c}{(mJy)} & (\hrs \ramin \rasec) & 
(\degree \arcmin \arcsec~) & (mas$^2$) & (deg) & Note \\
 (1) & (2) & (3) & (4) & (5) & \multicolumn{2}{c}{(6)} & 
\multicolumn{2}{c}{(7)} & \multicolumn{2}{c}{(8)} & (9) & (10) & (11) & (12) & (13) \\

\hline

VLBA Obs. & & & & & \multicolumn{2}{c}{} & \multicolumn{2}{c}{} &
\multicolumn{2}{c}{} & & & & & \\

\hline

NGC\,410  & 5 & 180 & 0.08 & 3.6$\times$1.9 & \multicolumn{2}{r}{$-$7.91} & \multicolumn{2}{r}{$<$0.4} & \multicolumn{2}{r}{$<$0.4} & \ldots & \ldots & \ldots  & \ldots & 4,c \\   

NGC\,524  & 5 & 180 & 0.08 & 4.5$\times$2.0 & \multicolumn{2}{r}{$-$9.89} &  \multicolumn{2}{r}{0.9} & \multicolumn{2}{r}{1.5} & 01
24 47.746 & +09 32 20.14 & 2.8$\times$1.8 & 168.02 & 4,c \\

NGC\,660  & 5 & 180 & 0.09 & 4.4$\times$1.9 & \multicolumn{2}{r}{$-$8.73} & \multicolumn{2}{r}{$<$0.5} & \multicolumn{2}{r}{$<$0.5} & \ldots & \ldots & \ldots & \ldots & 4,d \\

NGC\,1055 & 5 & 180 & 0.08 & 5.0$\times$1.9 & \multicolumn{2}{r}{$-$16.32} & \multicolumn{2}{r}{$<$0.4} & \multicolumn{2}{r}{$<$0.4} & \ldots & \ldots & \ldots & \ldots & 4,c \\

NGC\,1161 & 5 & 180 & 0.08 & 3.2$\times$1.8 & \multicolumn{2}{r}{$-$10.47} & \multicolumn{2}{r}{$<$0.5} & \multicolumn{2}{r}{$<$0.5} & \ldots & \ldots & \ldots & \ldots & 4,d \\

NGC\,3245 & 5 & 172 & 0.09 & 3.8$\times$1.7 & \multicolumn{2}{r}{$-$4.70} & \multicolumn{2}{r}{$<$0.5} & \multicolumn{2}{r}{$<$0.5} & \ldots & \ldots & \ldots & \ldots & 4,e \\

NGC\,3627 & 5 & 96 & 0.09 & 3.6$\times$1.4 & \multicolumn{2}{r}{$-$8.62} & \multicolumn{2}{r}{$<$0.3} & \multicolumn{2}{r}{$<$0.3} & \ldots & \ldots & \ldots & \ldots & 4,f \\

NGC\,4216 & 5 & 168 & 0.08 & 4.5$\times$1.8 & \multicolumn{2}{r}{$-$9.15} & \multicolumn{2}{r}{$<$0.4} & \multicolumn{2}{r}{$<$0.4} & \ldots & \ldots & \ldots & \ldots & 4,g \\

NGC\,4419 & 5 & 168 & 0.08 & 4.4$\times$1.8 & \multicolumn{2}{r}{$-$8.87} & \multicolumn{2}{r}{$<$0.4} & \multicolumn{2}{r}{$<$0.4} & \ldots & \ldots & \ldots & \ldots & 4,g \\

NGC\,4527 & 5 & 156 & 0.09 & 4.6$\times$1.6 & \multicolumn{2}{r}{$-$12.64} & \multicolumn{2}{r}{$<$0.5} & \multicolumn{2}{r}{$<$0.5} & \ldots & \ldots & \ldots & \ldots & 4,e \\

NGC\,4552 & 5 & 52 & 0.20 & 2.9$\times$1.2 & \multicolumn{2}{r}{2.40} & \multicolumn{2}{r}{99.5} & \multicolumn{2}{r}{99.5}& 12 35 39.807 & +12 33 22.83 & \ldots & \ldots & 2  \\



NGC\,5354 & 2.3 & 40 & 0.17 & 7.6$\times$3.8 &
\multicolumn{2}{r}{$-$3.44} &  \multicolumn{2}{r}{4.5} &
 \multicolumn{2}{r}{8.7} & 13 53 26.713 & +40 18 09.93 & 5.7$\times$3.2 & 102.02 & 4,g \\

          & 5 & 105 & 0.11 & 2.3$\times$1.6 &
          \multicolumn{2}{r}{6.74} &  \multicolumn{2}{r}{2.4} &
          \multicolumn{2}{r}{6.6} & 13
            53 26.712 & +40 18 09.94 & 3.9$\times$0.7 &  90.95  & 4,f
            \\

          & 5 & 48 & 0.14 & 3.5$\times$1.7 &
          \multicolumn{2}{r}{$-$8.05}&  \multicolumn{2}{r}{4.5} &
          \multicolumn{2}{r}{8.6} & 13 53 26.712 & +40 18 09.93 & 2.5$\times$1.5 & 103.11 & 4,g \\

          & 15 & 36 & 0.25 & 1.2$\times$0.6 &
          \multicolumn{2}{r}{$-$7.43} & \multicolumn{2}{r}{$<$1.3} &
          \multicolumn{2}{r}{$<$1.3} & \ldots & \ldots & \ldots & \ldots & 4,g \\

NGC\,5838 & 5 & 96 & 0.11 &  3.4$\times$1.4 &
\multicolumn{2}{r}{$-$1.26} & \multicolumn{2}{r}{$<$0.3} &
\multicolumn{2}{r}{$<$0.3} & \ldots & \ldots & \ldots & \ldots  & 4,f \\

NGC\,5846A  & 2.3 & 48 & 0.16 & 9.2$\times$3.7 &
\multicolumn{2}{r}{$-$0.51} &  \multicolumn{2}{r}{0.8}
& \multicolumn{2}{r}{2.8} & 15 06 29.292 & +01 36 20.35 & 9.0$\times$7.8 & 173.92 & 4,e \\
          
NGC\,5846A  & 5 & 112 & 0.05 & 3.4$\times$1.4 &
\multicolumn{2}{r}{$-$0.71} &  \multicolumn{2}{r}{0.6} &
 \multicolumn{2}{r}{1.1} & 15 06 29.292 & +01 36 20.34 & 2.8$\times$1.3 & 7.85 & 4,f \\ 

NGC\,5846A           & 5 & 48 & 0.12 & 4.7$\times$1.8 &
\multicolumn{2}{r}{$-$6.65} & \multicolumn{2}{r}{1.2} &
\multicolumn{2}{r}{1.5} & 15 06 29.292 & +01 36 20.34 & 2.3$\times$2.3 &
\ldots & 4,e \\

NGC\,5846B           & 5 & 112 & 0.05 & 3.4$\times$1.4 &
\multicolumn{2}{r}{$-$0.71} & \multicolumn{2}{r}{0.3} &
\multicolumn{2}{r}{0.6} & 15 06 29.292 & +01 36 20.44 & 4.5$\times$1.2 & 175.27 & 4,f \\

NGC\,5846C           & 5 & 112 & 0.05 & 3.4$\times$1.4 &
\multicolumn{2}{r}{$-$0.71} & \multicolumn{2}{r}{0.3} &
\multicolumn{2}{r}{0.7} & 15 06 29.292 & +01 36 20.45 & 5.2$\times$1.7 & 147.98 & 4,f \\

NGC\,5846D           & 15 & 48 & 0.16 & 1.7$\times$0.7 &
\multicolumn{2}{r}{$-$9.44} & \multicolumn{2}{r}{1.4} &
\multicolumn{2}{r}{1.9} & 15 06 29.292 & +01 36 20.40 & 1.3$\times$0.2 & 158.17 & 3,e \\
  
NGC\,5866 & 5 & 45 & 0.20 & \ldots & \multicolumn{2}{r}{\ldots} &
\multicolumn{2}{r}{7.0} & \multicolumn{2}{r}{8.4} &
15 06 29.499 & +55 45 47.57 & \ldots & (11) & 6  \\

\hline

\end{tabular}

\smallskip

{\sc References} -- (1) Filho, Barthel \& Ho 2002 from calibrated dated supplied by Dr. Jim Ulvestad; (2) Nagar \etal 2002; (3) Nagar \etal 2000;  (4)
this paper; (5) Cowan,
Romanishin \& Branch 1994; (6) Falcke \etal 2000. 

\smallskip

{\sc Observations} --  (a)   1995  July  13;  (b)  1999
Sept. 05; (c) 2001 Sept.  01; (d) 2001 Sept. 17; (e) 2001 Oct. 06; (f)
2000 June 22; (g) 2001 Sept. 08.
 
\end{minipage}

}

\rotatebox{90}{\usebox{\rotbox}}

\end{center}

\end{table*}

\normalsize

\clearpage



\begin{table*}

\scriptsize

\begin{center}

\sbox{\rotbox}{

\setcounter{table}{3}

\begin{minipage}{217mm} 

\caption{{\bf Summary of the AGN candidate properties (published, archival or
    new).}
Col. 1: Source name.
Col. 2: Adopted distance for H$_{\rm 0}$=75 km s$^{-1}$
Mpc$^{-1}$ (Tully 1988 or Ho, Filippenko \& Sargent 1997a).
Col. 3--5: Extinction-corrected [OI]$\lambda$6300, H$\alpha$, and
H$\beta$ line luminosity (Ho, Filippenko \& Sargent 1997a, 2003).
Col. 6: Published or new high resolution VLBA 5\,GHz (or VLA; see footnote) radio core luminosity.  When multiple 5\,GHz, VLBA
observations are available, the longest integration
time observations is quoted.
Col. 7: Archival and/or published {\it Chandra} hard X-ray (2-10 keV) nuclear luminosity.
Col. 8: Nuclear B-band magnitude.
Col. 9: Logarithm of the radio to nuclear B band ratio.
Col. 10: Logarithm of the radio to hard X-ray ratio.
Col. 11: Logarithm of the X-ray to H$\alpha$ ratio. 
Col. 12: Central velocity dispersion (Lyon/Meudon Extragalactic
Database or HYPERCAT  -- NGC\,660, 3627, 4419, 4527, and 5354).
Col. 13: Estimated black hole mass obtained from the velocity dispersion. 
Col. 14: Logarithm of the Eddington luminosity.
Col. 15: Logarithm of the X-ray to Eddingtion luminosity ratio.
}

\smallskip

\begin{tabular}{l c cc c c cc cc cc cc cc cc c c c cc}

\hline
\hline

 & $D$ & \multicolumn{2}{c}{log L([O~{\sc i}])} & log L(H$\alpha$) & log
 L(H$\beta$) & \multicolumn{2}{c}{log L$_{\rm R}$} &
 \multicolumn{2}{c}{log L$_{\rm X}$} & \multicolumn{2}{c}{M$_{\rm B}^{\rm
     nuc}$} &\multicolumn{2}{c}{}  & \multicolumn{2}{c}{} & \multicolumn{2}{c}{} & $\sigma_c$ & M$_{\rm BH}$ & log L$_{\rm Edd}$ & \multicolumn{2}{c}{} \\
Galaxy & (Mpc) & \multicolumn{2}{c}{(erg s$^{-1}$)} &
(erg s$^{-1}$) &  (erg s$^{-1}$) &  \multicolumn{2}{c}{(erg
  s$^{-1}$)} & \multicolumn{2}{c}{(erg s$^{-1}$)} & \multicolumn{2}{c}{(mag)} &
\multicolumn{2}{c}{$\frac{{\rm L_{\rm R}}}{{\rm L_{\rm o}}}$} & \multicolumn{2}{c}{$\frac{{\rm
      L_{\rm R}}}{{\rm L_{\rm X}}}$} &
\multicolumn{2}{c}{$\frac{\rm {L_{\rm X}}}{{\rm
      L(\rm H\alpha)}}$} & (km s$^{-1}$) & (M$_{\odot}$) & (erg s$^{-1}$) &
  \multicolumn{2}{c}{$\frac{\rm L_{X}}{\rm L_{Edd}}$} \\
 (1) & (2) & \multicolumn{2}{c}{(3)} & (4) & (5) & \multicolumn{2}{c}{(6)} &
 \multicolumn{2}{c}{(7)} & \multicolumn{2}{c}{(8)}  & \multicolumn{2}{c}{(9)} &
 \multicolumn{2}{c}{(10)} & \multicolumn{2}{c}{(11)} & (12) & (13) &  (14) & \multicolumn{2}{c}{(15)}   \\

\hline

N\,410 & 70.6 & \multicolumn{2}{r}{$<$38.4} & 39.43 & 39.46 &  \multicolumn{2}{r}{$<$37.08} & \multicolumn{2}{r}{\ldots} & \multicolumn{2}{r}{$-$12.82} & \multicolumn{2}{r}{$<$1.6} & \multicolumn{2}{r}{\ldots} & \multicolumn{2}{r}{$<$2.6} & 285 & 5.5$\times$10$^8$  & 46.85 & \multicolumn{2}{r}{\ldots}  \\

N\,524  & 67.0 & \multicolumn{2}{r}{$<$37.7} & 38.58 & 38.08 &  \multicolumn{2}{r}{37.61} & \multicolumn{2}{r}{\ldots} &
\multicolumn{2}{r}{$-$8.76} & \multicolumn{2}{r}{3.7} & \multicolumn{2}{r}{\ldots} & \multicolumn{2}{r}{\ldots} & 245 & 3.1$\times$10$^8$ & 46.60 & \multicolumn{2}{r}{\ldots}  \\

N\,660  & 11.8 &  \multicolumn{2}{r}{39.1} & 40.41 & 39.27 & \multicolumn{2}{r}{$<$35.62} &  \multicolumn{2}{r}{38.33} & \multicolumn{2}{r}{$-$12.26} & \multicolumn{2}{r}{$<$0.3} & \multicolumn{2}{r}{$<-$2.7} & \multicolumn{2}{r}{$<-$2.1} & 128 & 2.2$\times$10$^7$ & 45.46 & \multicolumn{2}{r}{$-$7.1} \\

N\,1055  & 12.6 &  \multicolumn{2}{r}{36.7} & 37.92 & \ldots & \multicolumn{2}{r}{$<$35.58} & \multicolumn{2}{r}{$<$38.39} &
\multicolumn{2}{r}{\ldots} & \multicolumn{2}{r}{\ldots} & \multicolumn{2}{r}{\ldots} & \multicolumn{2}{r}{$<$0.5} & 80 & 3.4$\times$10$^6$ & 44.65 &  \multicolumn{2}{r}{$<-$6.3} \\

N\,1161 & 25.9 & \multicolumn{2}{r}{$<$38.2} & 39.01 & 38.39 & \multicolumn{2}{r}{$<$36.31} & \multicolumn{2}{r}{\ldots} &
\multicolumn{2}{r}{$-$9.68}  & \multicolumn{2}{r}{$<$2.1} & \multicolumn{2}{r}{\ldots} & \multicolumn{2}{r}{$<$2.3} & 285 & 5.6$\times$10$^8$ & 46.86  & \multicolumn{2}{r}{\ldots} \\

N\,3245 & 22.2 &  \multicolumn{2}{r}{39.0} & 40.03 & 39.35 & \multicolumn{2}{r}{$<$36.17} & \multicolumn{2}{r}{38.89} & \multicolumn{2}{r}{$-$12.50} & \multicolumn{2}{r}{$<$0.8} & \multicolumn{2}{r}{$<-$2.7} & \multicolumn{2}{r}{$<-$1.1} & 211 & 1.7$\times$10$^8$ & 46.34 & \multicolumn{2}{r}{$-$7.5} \\

N\,3627 & 6.6 &  \multicolumn{2}{r}{38.3}  &  39.16 & 38.39 & \multicolumn{2}{r}{$<$34.90} & \multicolumn{2}{r}{\lax38.06} &\multicolumn{2}{r}{$-$9.68} & \multicolumn{2}{r}{$<$0.7} & \multicolumn{2}{r}{$\sim-$3.2} & \multicolumn{2}{r}{$<-$1.1} & 124 & 2.0$\times$10$^7$ & 45.44 & \multicolumn{2}{r}{\lax$-$7.4} \\

N\,4216 & 16.8 &  \multicolumn{2}{r}{38.0} & 38.89 & 38.25 & \multicolumn{2}{r}{$<$35.83} & \multicolumn{2}{r}{\ldots} &
\multicolumn{2}{r}{$-$9.26} & \multicolumn{2}{r}{$<$1.8} & \multicolumn{2}{r}{\ldots} & \multicolumn{2}{r}{\ldots} & 219 & 1.9$\times$10$^8$ & 46.39 & \multicolumn{2}{r}{\ldots} \\

N\,4419 & 16.8 &  \multicolumn{2}{r}{38.1} & 40.72 & 39.61 &
\multicolumn{2}{r}{$<$35.83} & \multicolumn{2}{r}{\ldots} &
\multicolumn{2}{r}{$-$13.26} & \multicolumn{2}{r}{$<$0.2} &
\multicolumn{2}{r}{\ldots} & \multicolumn{2}{r}{\ldots} & 101 &
8.7$\times$10$^6$ & 45.05 & \multicolumn{2}{r}{\ldots} \\

N\,4527 & 13.5 &  \multicolumn{2}{r}{39.0} & 40.12 & 39.12 &
\multicolumn{2}{r}{$<$35.74} & \multicolumn{2}{r}{\ldots} &
\multicolumn{2}{r}{$-$11.82} & \multicolumn{2}{r}{$<$0.6} & \multicolumn{2}{r}{\ldots} & \multicolumn{2}{r}{\ldots} & 214 & 1.8$\times$10$^8$ & 46.37 & \multicolumn{2}{r}{\ldots} \\

N\,4552$^{a}$ & 16.8 & \multicolumn{2}{r}{$<$37.7} & 38.55 & 38.05 &  \multicolumn{2}{r}{38.23} &  \multicolumn{2}{r}{39.41} & \multicolumn{2}{r}{$-$8.68} & \multicolumn{2}{r}{4.4} & \multicolumn{2}{r}{$-$1.2} & \multicolumn{2}{r}{0.9} & 264 & 4.1$\times$10$^8$ & 46.73 & \multicolumn{2}{r}{$-$7.3} \\

N\,5354 & 32.8 &  \multicolumn{2}{r}{37.9} &  38.71 & 38.26 &  \multicolumn{2}{r}{37.63} & \multicolumn{2}{r}{\ldots} & \multicolumn{2}{r}{$-$9.29} & \multicolumn{2}{r}{3.5} & \multicolumn{2}{r}{\ldots} & \multicolumn{2}{r}{\ldots} & 221 & 2.0$\times$10$^8$ & 46.41 & \multicolumn{2}{r}{\ldots}  \\

N\,5838 & 28.5 & \multicolumn{2}{r}{$>$37.7} & \ldots & \ldots & \multicolumn{2}{r}{$<$36.17} & \multicolumn{2}{r}{\ldots} &
\multicolumn{2}{r}{\ldots} & \multicolumn{2}{r}{\ldots} &
\multicolumn{2}{r}{\ldots} & \multicolumn{2}{r}{\ldots} & 297 &
6.6$\times$10$^8$ & 46.93 & \multicolumn{2}{r}{\ldots} \\

N\,5846$^{b,c}$ & 28.5 &  \multicolumn{2}{r}{38.0} & 39.01 & 38.43 &  \multicolumn{2}{r}{36.73} &  \multicolumn{2}{r}{38.37} & \multicolumn{2}{r}{$-$9.79} & \multicolumn{2}{r}{2.4} & \multicolumn{2}{r}{$-$1.6} & \multicolumn{2}{r}{$-$0.7} & 250 & 3.3$\times$10$^8$ & 46.63 & \multicolumn{2}{r}{$-$8.3}  \\

N\,5866$^{d}$ & 15.3 & \multicolumn{2}{r}{$<$37.9} & 38.82 & 37.99 &
\multicolumn{2}{r}{37.07} & \multicolumn{2}{r}{$<$38.26} &
\multicolumn{2}{r}{$-$8.50} & \multicolumn{2}{r}{3.3} & \multicolumn{2}{r}{$>-$1.2} & \multicolumn{2}{r}{$-$0.6} & 142 & 3.4$\times$10$^7$ & 45.65 & \multicolumn{2}{r}{$<-$7.4} \\

N\,7331$^{e}$ & 14.3 & \multicolumn{2}{r}{$<$37.7} & 38.73 & 38.14 &
\multicolumn{2}{r}{$<$35.17} &  \multicolumn{2}{r}{38.45} &
\multicolumn{2}{r}{$-$8.94} & \multicolumn{2}{r}{$<$1.2} & \multicolumn{2}{r}{$<-$3.3} & \multicolumn{2}{r}{$-$0.3} & 146 & 3.8$\times$10$^7$ & 45.69 & \multicolumn{2}{r}{$-$7.2}  \\ 
   
\hline

\end{tabular}

$^{\it a}$ VLBA, 5\,GHz, 2~mas resolution radio core luminosity from
Nagar \etal (2002).

$^{\it b}$ VLBA, 5\,GHz, 2~mas resolution radio core luminosity
  given for the hypothetical core -- component A.

$^{\it c}$ Hard X-ray (2-10 keV) luminosity from Terashima \& Wilson (2003).

$^{\it d}$ VLBA, 5\,GHz, 2~mas resolution radio core luminosity
from Falcke \etal (2002).

$^{\it e}$ VLA, 5\,GHz, 2\arcsec~resolution radio core
  luminosity from Cowan, Romanishin \& Branch (1994).

\end{minipage}

}

\rotatebox{90}{\usebox{\rotbox}}

\end{center}

\end{table*}

\normalsize

\clearpage



\setcounter{figure}{0}
\begin{figure*}
\leavevmode
\centerline{
\epsfxsize=6.5cm
\epsffile{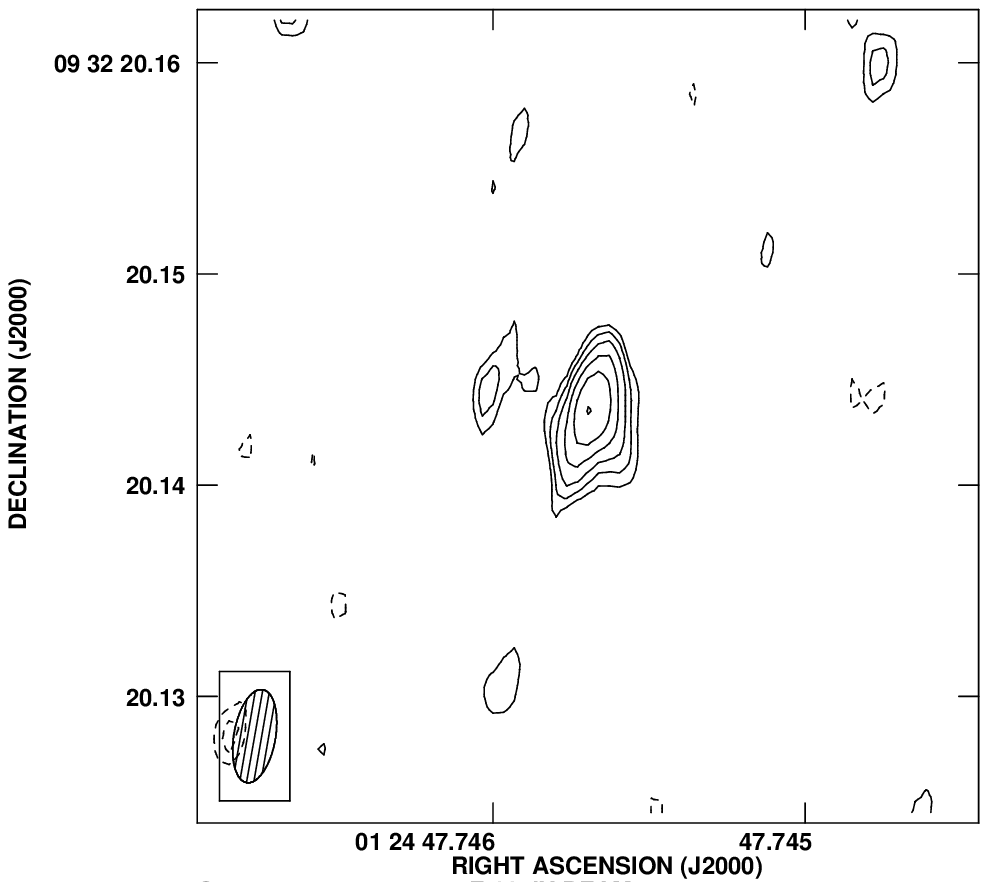}
(a)
}
\end{figure*}

\begin{figure*}
\leavevmode
\centerline{
\epsfxsize=6.5cm
\epsffile{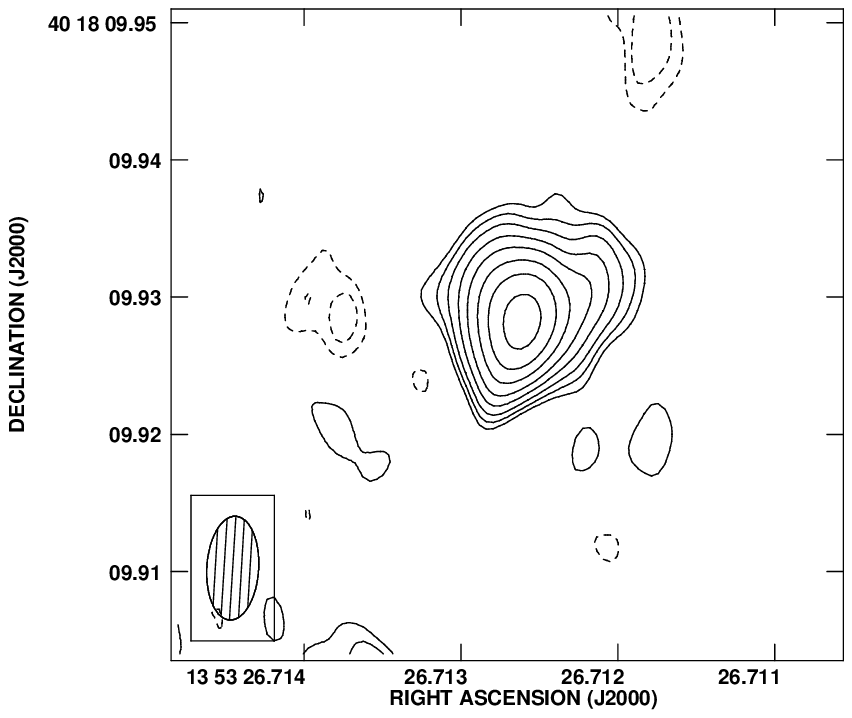}
(b)
\epsfxsize=6.5cm
\raisebox{0.3cm}{
\epsffile{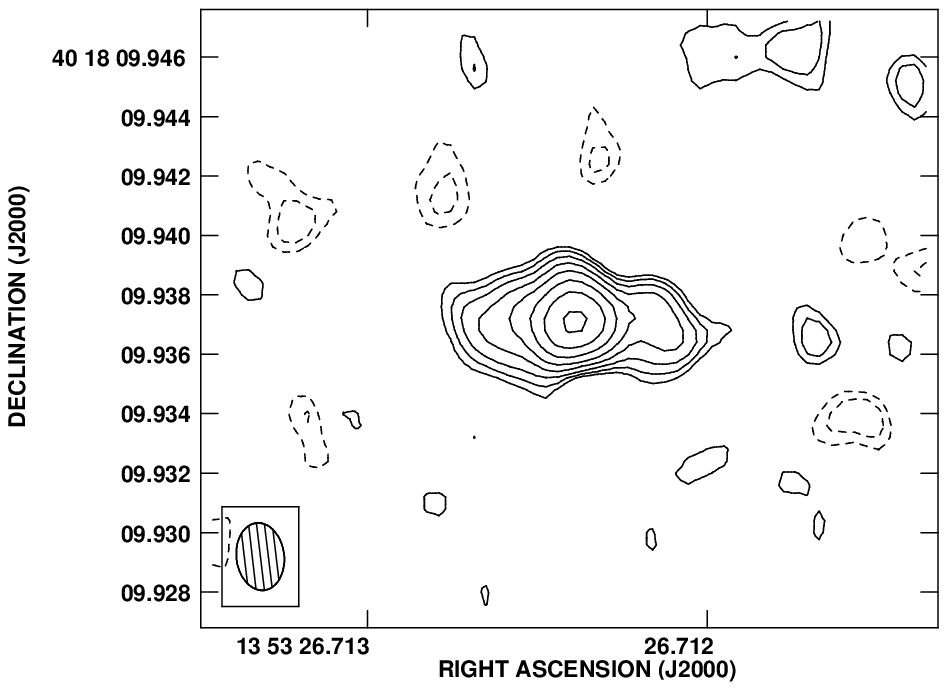}
(c)
}
}
\caption{Radio emission contours with the following multiples of the
  $rms$ noise (Table~1):  ($-$3, $-$2.1, 2.1, 3, 4.2, 6, 8.4, 12,
  16.8, 24). 
The size of the restoring beam in milliarcseconds, observing epoch,
integration time in minutes, and frequency in GHz are given in parentheses after
each object name (see Table~1).  {\it  (a)}  NGC\,524
(4.5 $\times$ 2.0; 2001 Sept. 01; 180; 5),   {\it  (b)}  NGC\,5354
(7.6 $\times$ 3.8; 2001 Sept. 17; 40; 2.3), {\it  (c)}  NGC\,5354
(2.3 $\times$ 1.6; 200 June 22; 105; 5).}

\end{figure*}

\clearpage


\setcounter{figure}{1}
\begin{figure*}
\leavevmode
\centerline{
\epsfxsize=6.5cm
\raisebox{2.5cm}{
\epsffile{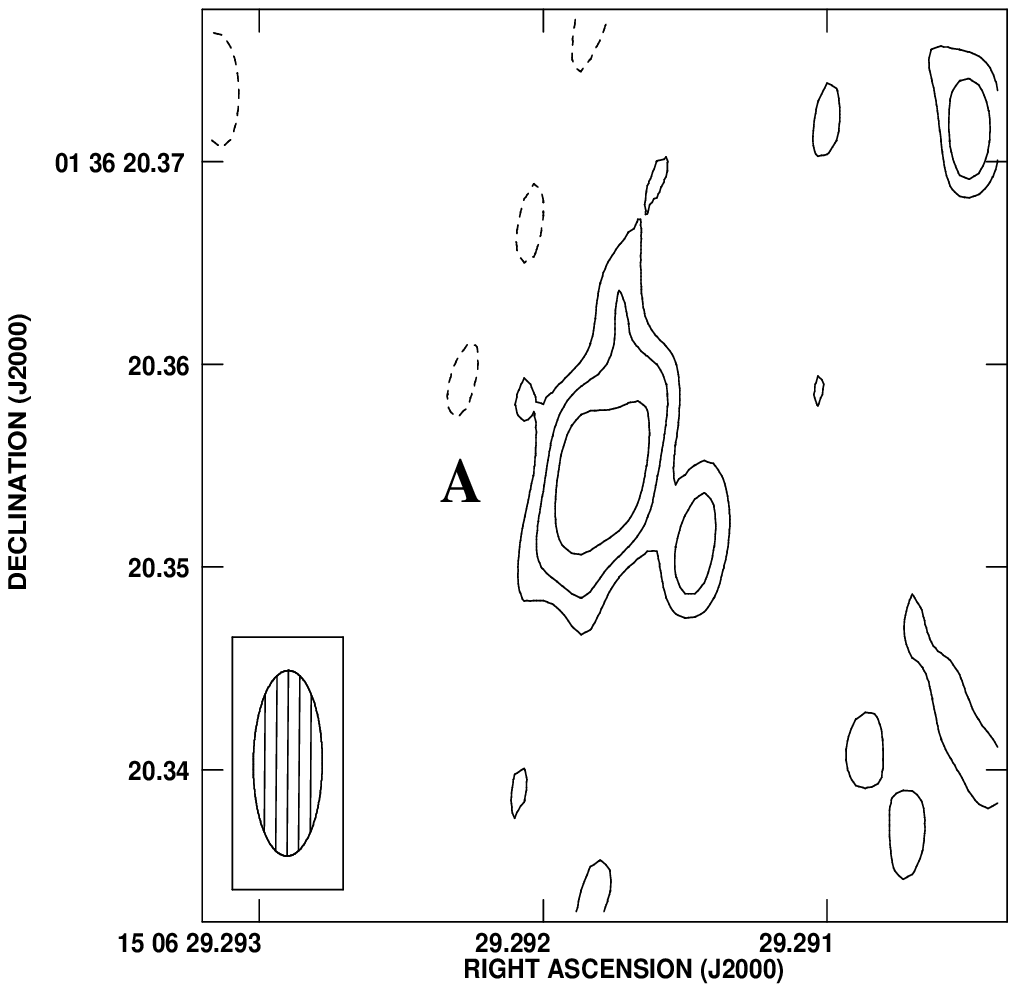}
(a)
}
\epsfxsize=5.0cm
\epsffile{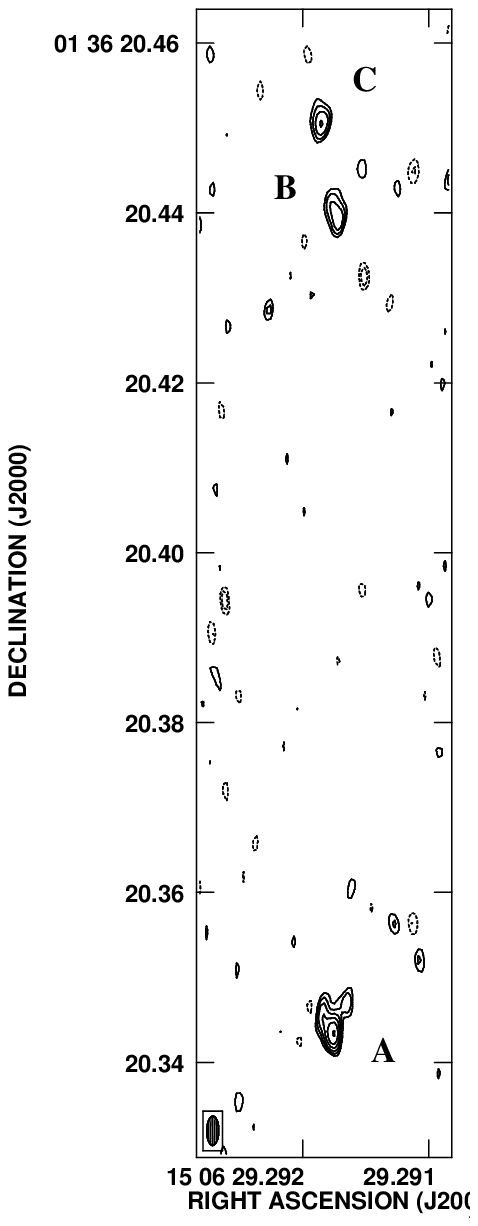}
(b)
}
\end{figure*}

\begin{figure*}
\leavevmode
\centerline{
\epsfxsize=6.5cm
\epsffile{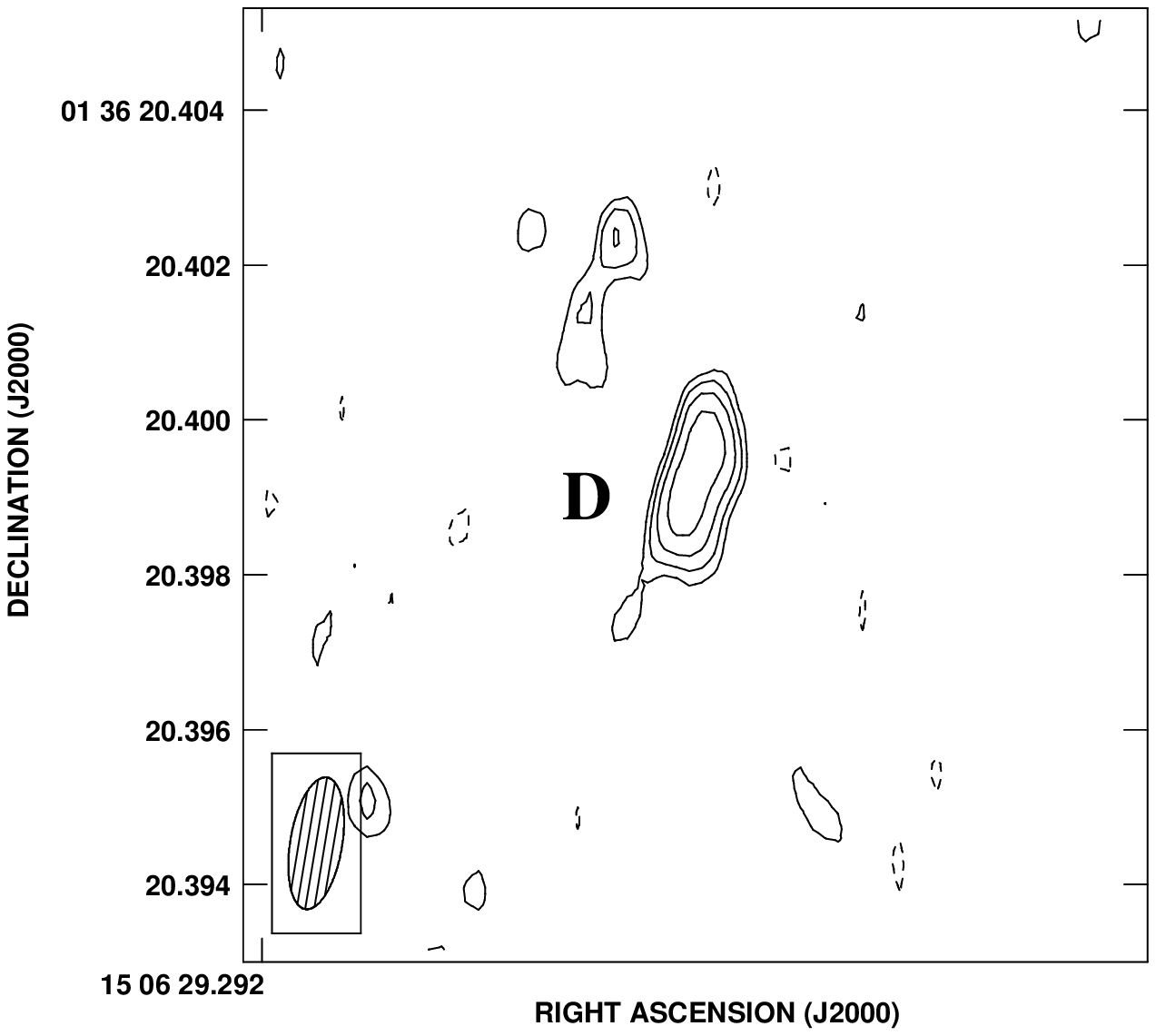}
(c)
}
\caption{Same as Figure~1.  {\it  (a)}  NGC\,5846
(9.2 $\times$ 3.7; 2001 Oct. 06; 48; 2.3), {\it (b)} NGC\,5846
(3.1 $\times$ 1.4; 2000 June 22; 112; 5), {\it  (c)}  NGC\,5846   (1.7 $\times$ 0.7; 2001
Oct. 06; 48; 15).}

\end{figure*}

\end{document}